\documentclass[twocolumn,preprintnumbers,superscriptaddress]{revtex4-2}

\usepackage{blindtext}
\usepackage{amsmath}
\usepackage{amsfonts}
\usepackage{graphicx}
\usepackage{float}
\usepackage{booktabs}
\usepackage{gensymb}
\usepackage{hyperref}
\usepackage{braket}
\usepackage{amssymb}
\usepackage{array}
\usepackage{soul}
\usepackage{algorithm}
\usepackage{algpseudocode}
\usepackage{tabularx}
\usepackage[table,xcdraw]{xcolor}
\usepackage{enumitem}
\usepackage{listings}

\definecolor{codegreen}{rgb}{0,0.6,0}
\definecolor{codegray}{rgb}{0.5,0.5,0.5}
\definecolor{codepurple}{rgb}{0.58,0,0.82}
\definecolor{backcolour}{rgb}{0.97,0.97,0.97}

\definecolor{blue}{rgb}{0,0,0.9}
\definecolor{orange}{rgb}{1,0.6,0}
\definecolor{red}{rgb}{1,0,0}
\definecolor{pink}{rgb}{1,0.41,0.71}

\usepackage{color}
\definecolor{blue}{rgb}{0,0,0.9}
\definecolor{orange}{rgb}{1,0.3,0}
\definecolor{darkblue}{rgb}{0,0,0.5}
\definecolor{darkgreen}{rgb}{0.0, 0.7, 0.0}

\hypersetup{
    bookmarksnumbered=true,
    unicode=false,
    pdfstartview={FitH},
    pdftitle={},
    pdfauthor={},
    pdfsubject={},
    pdfcreator={},
    pdfproducer={},
    pdfkeywords={},
    pdfnewwindow=true,
    colorlinks=true,
    linkcolor=darkblue,
    citecolor=darkblue,
    filecolor=darkblue,
    urlcolor=darkblue
}

\definecolor{jycolor}{rgb}{0.913,0.588,0.478}

\definecolor{amcolor}{rgb}{0.7,0.64,0.1}

\newcommand{\IHPC}{Institute of High Performance Computing (IHPC), Agency for Science, Technology and Research (A*STAR), 1 Fusionopolis Way, \#16-16 Connexis, Singapore 138632, Republic of Singapore}
\newcommand{\TIFlab}{TIF Lab, Dipartimento di Fisica, Università degli Studi di Milano, Milan, Italy}
\newcommand{\INFN}{INFN, Sezione di Milano, I-20133 Milan, Italy}
\newcommand{\CERN}{Theoretical Physics Department, CERN, CH-1211 Geneva 23, Switzerland}
\newcommand{\TII}{Quantum Research Center, Technology Innovation Institute, Abu Dhabi, United Arab Emirates}



\begin{document}
\date{\today}

\title{Designing a Machine Learning-Driven, Cross-Hardware Emulator for Noisy Quantum Computers with Gate-Based Protocols}

\author{Matthew Ho}
\email{matthew\_ho@a-star.edu.sg}
\affiliation{\IHPC}
\thanks{MH and JYK contributed equally}

\author{Jun Yong Khoo}
\email{khoojy@a-star.edu.sg}
\affiliation{\IHPC}
\thanks{MH and JYK contributed equally}

\author{Adrian M. Mak}
\affiliation{\IHPC}

\author{Stefano Carrazza}
\affiliation{\TIFlab}
\affiliation{\INFN}
\affiliation{\CERN}
\affiliation{\TII}

\begin{abstract}
    Quantum computer emulators model the behavior and error rates of specific quantum processors. Without accurate noise models in these emulators, it is challenging for users to optimize and debug executable quantum programs prior to running them on the quantum computer, as device-specific noise is not properly accounted for.
    To overcome this challenge, we design a machine learning (ML)-driven approach to construct approximate device-specific emulators that applies to different hardware platforms.
    We apply supervised ML on a pre-generated library containing simulated gate set tomography training data.
    The ML model then analyses gate set tomography data from a target quantum computer to predict its noise model, which is in turn used to construct the device-specific emulator.
    We demonstrate the effectiveness of our protocol’s emulator in estimating the unitary coupled cluster energy of the H$_2$ molecule and compare the results with those from actual quantum hardware. 
    Remarkably, our noise model captures device noise with high accuracy, achieving a percentage relative error of just 0.128\% in expectation value relative to the actual quantum hardware.
    Importantly, we show that even without access to pulse-level control, noise from the quantum computer can nonetheless be characterized and independently validated by our protocol.
\end{abstract}

\maketitle

\section{Introduction}

In the current noisy intermediate-scale quantum~\cite{Preskill2018quantumcomputing} (NISQ) computing era, access to quantum computers are predominantly cloud-based~\cite{AmazonBraket, IBMQuantum, quantinuum, chen2024benchmarking, wurtz2023aquila, boothby2020next}.
Despite the host of available NISQ computers from different service providers, their accessibility is often limited or costly; users can be subjected to long queuing times or have to pay for circuit executions.
Costs will increase when compute-intensive problems require many iterations, especially when using quantum algorithms that are variational in nature~\cite{peruzzo2014variational, farhi2014quantum, cerezo2021variational, benedetti2019pqc, qi2026tensorhyper}. As a result, despite variational quantum algorithms being the most prevalent type of quantum algorithm for NISQ computers~\cite{cerezo2021variational, zeng2021simulating, lau2022nisq, du2022architecture}, with the exception of simple toy problems, it is unfeasible for most users to perform the definitive routine of quantum circuit parameter optimization directly on NISQ computers. 
Instead, parameter optimization is often done via classical simulations, and the optimized variational quantum circuit is then run on the NISQ device for validation or comparison~\cite{umer2024nonlinear, chee2024omp2, per2025chemically}.
In addition, noise-induced effects such as barren plateaus further complicate optimization in variational quantum algorithms \cite{wang2021barren, larocca2025barren}.

To maximize the efficiency of the validation run on NISQ computers, it is crucial to have classical simulators that can faithfully replicate the behavior of individual NISQ devices through device-specific emulators. 
Users can construct an emulator for their chosen quantum computer, but not all input parameters required for a comprehensive noise model are regularly updated or readily available from the hardware provider. 
For instance, IQM And Rigetti do not include one-qubit and two-qubit gate durations for their devices in their calibration data on Amazon Braket, which are required as inputs to noise models that incorporate the thermal relaxation channel~\cite{georgopoulos2021modeling}.
These parameters are derived from calibration protocols that require access to pulse-level control that the general user lacks.
However, even in the case of IBMQ in which device-specific emulators are provided based on realistic noise models, they often underestimate the effect of actual hardware noise~\cite{chee2024omp2, bravo2024methodology}.
In the absence of device-specific emulators, users typically rely on noiseless state-vector simulations of their quantum algorithms, the results of which would substantially deviate from those obtained on currently available NISQ devices. 
The lack of such device-specific emulators makes job submissions to NISQ devices costly and time-consuming, with limited conclusions drawn from the results. 
More broadly, machine learning techniques have emerged as a promising paradigm for extracting useful information from quantum systems and noisy quantum data~\cite{cerezo2022qml, huang2021power}.

Without the latest calibration data, general users can only perform gate-based characterization protocols such as randomized benchmarking~\cite{knill2008randomized, magesan2012characterizing} and gate set tomography (GST)~\cite{greenbaum2015GST, blume2017GST, nielsen2021GST}. 
Conventionally, these protocols only enable a partial noise profiling of the NISQ device;
all other device characteristics needed for a more complete and up-to-date noise model such as the qubit relaxation time from the $\ket{1}$ to $\ket{0}$ state (T1 time), the time that qubits preserve coherence (T2 time), as well as calibrated native gate times cannot be obtained. 
Moreover, it is impractical for a resource-limited user to perform certain gate-based protocols such as process tomography on every native gate.

Recognizing the need for realistic emulators of hardware, several approaches have been proposed to model device noise. 
Prior work has explored parameter optimization of thermal relaxation and depolarizing noise~\cite{georgopoulos2021modeling, bravo2024methodology}. Other methods apply noise through perturbations of the Hamiltonian for single-qubit gates, Gaussian phase noise for two-qubit gates, and Gaussian coherent noise to account for control errors~\cite{martin2020digital}. 
These approaches use an underlying physical model of the device as a starting point, such that the accuracy of the noise model is limited to that of the physical model as well as
the accessibility of precise device characterization data.
In particular, general users typically lack knowledge of in-situ device conditions of their QPU of interest, such that their implementation of the best available physical models inevitably omit or oversimplify these setup-specific noise effects.

Existing approaches to noise learning, such as Lindblad-based reconstruction~\cite{malekakhlagh2025efficient} and time-series methods~\cite{berg2025large}), typically rely on physics-informed models and/or time-resolved data. In contrast, while our approach is constructed from standard physically motivated noise channels, its parameter determination is entirely data-driven and operates using only gate-based GST data, without requiring pulse-level access or detailed calibration parameters. This makes the method more accessible to general users, albeit at the cost of reduced physical interpretability.

In this work, we propose a general GST-based protocol that leverages supervised machine learning (ML) to construct heuristic device noise models agnostic to device physics. 
The goal of our protocol is to closely approximate noise profiles of specific NISQ devices even when device-specific noise characterizations are unavailable.
By eliminating the need for device-specific knowledge—particularly setup-dependent details or information obtainable only through pulse-level control—our protocol, and consequently the noisy emulator, are accessible and directly implementable by general users.
To demonstrate how general users can adapt our protocol to construct emulators for the NISQ device of their choice, we apply it with a minimal composite noise model to two selected qubits of the IQM Garnet quantum processor accessed through Amazon Braket~\cite{iqm, abdurakhimov2024technology}.
We demonstrate that our protocol's emulator, based on a machine-learned heuristic model, has the potential to outperform one that requires device calibration data, thereby opening a complementary paradigm to realistic qubit characterization and noise emulation.

As we will discuss in detail in Sec.~\ref{sec:methods}, our protocol utilizes neural networks trained on GST data simulated over a range of noise parameter values to predict the noise parameter values that best fit GST data obtained from the target NISQ device.
In Sec.~\ref{sec:results}, we evaluate the performance of our protocol's emulator by benchmarking its results for a quantum chemistry task against results obtained from IQM Garnet.
Our benchmark shows that GST can detect the footprints from device noise, which can then be distilled by neural networks to fit into commonly used noise models. Our emulator is then constructed with a minimal heuristic noise model comprising depolarizing, amplitude damping, dephasing, and readout noise channels. It is capable of achieving simulation results that have appreciably small deviation from the actual noisy device.
To conclude, we discuss in Sec.~\ref{sec:conclusions} the impact of our work, its wide applicability and utility, as well as follow-up work and future directions.

\begin{figure*}
    \centering
    \includegraphics[width=1.0\linewidth]{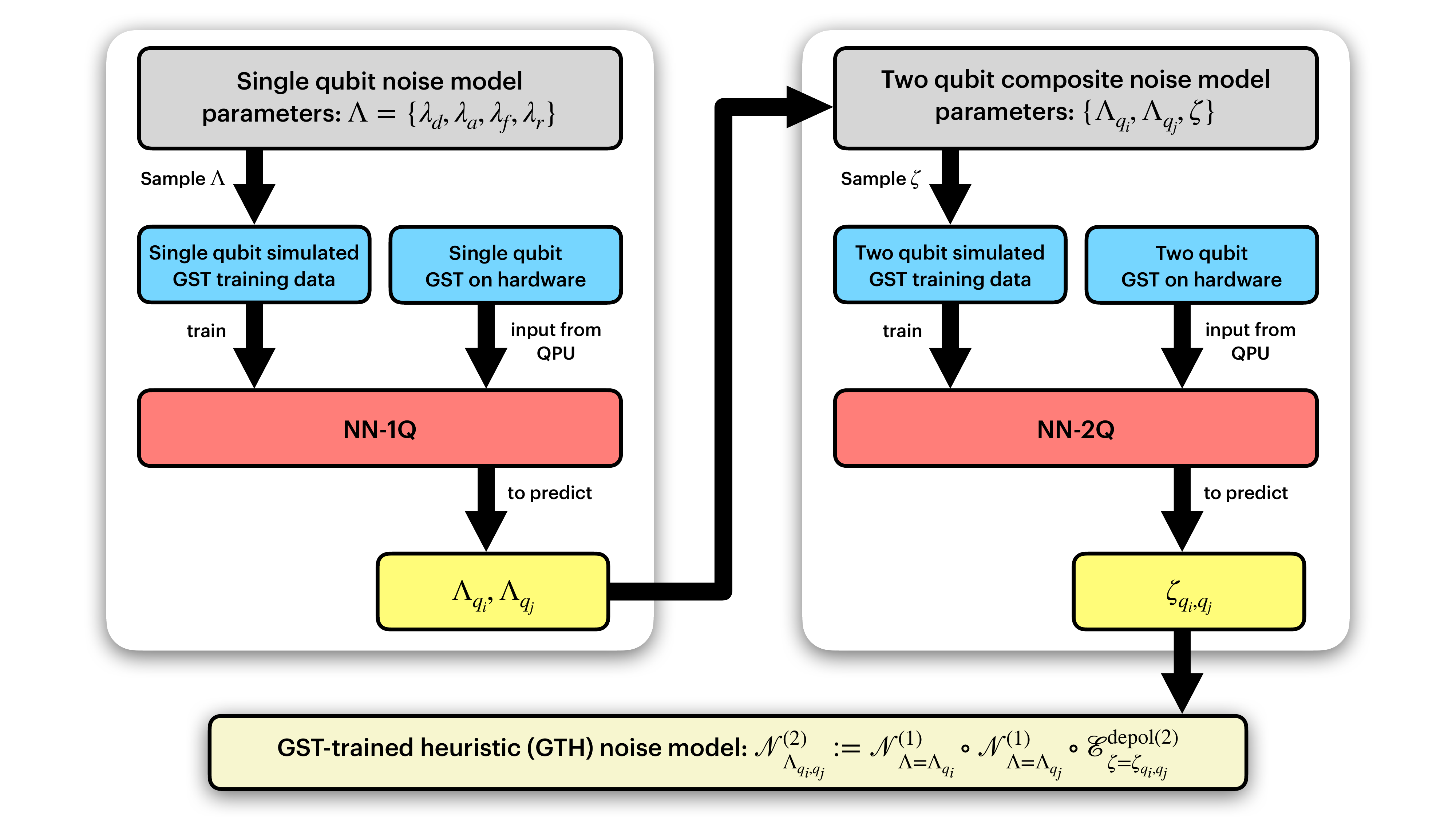}
    \caption{Schematic of the proposed ML-driven, generalized gate-based protocol for processing simulated and hardware GST data. A neural network for single-qubit noise (NN-1Q) predict noise parameters $\Lambda _{q_i}, \Lambda _{q_j}$, while a neural network for two-qubits (NN-2Q) predicts $\zeta_{q_i, q_j}$ for the target qubits $q_i$ and $q_j$. These predictions are used to construct the GST-trained heuristic (GTH) noise model $\mathcal{N}^{(2)}_{\Lambda_{q_i, q_j}}$.}
    \label{fig.workflow}
\end{figure*}

\section{Methods}~\label{sec:methods}

Our ML-driven protocol comprises three key steps: 
\begin{enumerate}
    \item choose a noise model with a set of input noise parameters that is capable of emulating QPU noise,
    \item design a gate-based protocol to acquire data from a target QPU,
    \item perform machine learning to predict the target QPU's noise model parameters based on its characterization data.
\end{enumerate}

In the respective subsections (Sec.~\ref{subsec:noisemodels}-\ref{sec.training_and_evaluating}), we concretize each step of our protocol to construct the GST-trained heuristic (GTH) noise model. We summarize the key steps of the protocol in the form of a recipe in Sec.~\ref{sec.recipe} and provide a corresponding schematic in Fig.~\ref{fig.workflow}.

\subsection{Choose a noise model}~\label{subsec:noisemodels}

In the first step of our protocol, we define a general $n$-qubit noise model comprising $l$-sequential applications of distinct noise channels $\mathcal{E}_{\lambda_j}$, each parametrized by $\lambda_j$,
\begin{equation}
\mathcal{N}^{(n)}_{\Lambda} 
    = \mathcal{E}_{\lambda_1} \circ \mathcal{E}_{\lambda_2} \circ ... \circ \mathcal{E}_{\lambda_l}.
\end{equation}
Here, $\Lambda = \{\lambda_1, \lambda_2, ..., \lambda_l \}$ denotes the set of input parameters $\lambda_j$ corresponding to each of the noise channels.
The action of each noise channel on the density matrix $\rho$ of a system is 
$\mathcal{E}_{\lambda_j}(\rho) = \sum_{m=1}^{k} K_{\lambda_j,m} \rho K_{\lambda_j,m}^{\dagger}$ with Kraus operators $K_{\lambda_j,m}$  that are constrained by $\sum_{m=1}^{k} K_{\lambda_j,m}^{\dagger} K_{\lambda_j,m} = I$. 
We denote a single qubit noise model acting on qubit $q_i$ as $\mathcal{N}^{(1)}_{\Lambda_{q_i}}$ and a two qubit noise model acting on qubits $q_i$ and $q_j$ as $\mathcal{N}^{(2)}_{\Lambda_{q_i, q_j}}$.

In this work, we focus on heuristic noise models with the sole objective of replicating as close as possible (within statistical error due to shot noise) the results of a target QPU.
While our protocol is compatible with any arbitrary noise model, the ability to capture the effects of QPU noise depends on the complexity of the heuristic noise model through the noise channels it includes.
More complex noise models would potentially enable an even more accurate emulation of the device noise at the cost of having a more challenging machine learning task; noise models that are too simple would likely suffer from the lack of basis channels and struggle to find a close fit to the hardware data such as that obtained from GST discussed in the next section.
To keep the difficulty of the machine learning task under control, we consider a minimally complex heuristic noise model comprising one-qubit depolarizing noise, amplitude damping, dephasing, and reaodut noise channels, respectively parametrized by 
$\Lambda = \{\lambda_d,\lambda_a, \lambda_f, \lambda_r\}$,
\begin{equation} \label{eq.noise_model_1qb}
    \mathcal{N}^{(1)}_{\Lambda} := \mathcal{E}^{\text{depol}(1)}_{\lambda_d} \circ \mathcal{E}^{\text{ampli}(1)}_{\lambda_a} \circ \mathcal{E}^{\text{deph}(1)}_{\lambda_f} \circ \mathcal{E}^{\text{readout}(1)}_{\lambda_r},
\end{equation}
and an additional two-qubit depolarizing noise channel parametrized by a single parameter $\zeta$,
\begin{equation}
    \mathcal{E}^{\text{depol}(2)}_{\zeta}. 
\end{equation}
Finally, we define the two-qubit composite noise on target qubits $q_i$ and $q_j$ as,
\begin{equation} \label{eq.noise_model_2qb}
    \mathcal{N}^{(2)}_{\Lambda_{q_i, q_j}} := \mathcal{N}^{(1)}_{\Lambda=\Lambda_{q_i}} \circ \mathcal{N}^{(1)}_{\Lambda=\Lambda_{q_j}} \circ \mathcal{E}_{\zeta=\zeta_{q_i, q_j}}^\text{depol(2)},
\end{equation}
parametrized by qubit specific parameters $\Lambda_{q_i, q_j} = \{\Lambda_{q_i}, \Lambda_{q_j}, \zeta_{q_i, q_j}\}.$

The noise channels we include are prototypical decoherence processes for two-level systems that are agnostic to the physical nature of the qubit, and are required to provide the minimal set of mechanisms to emulate the effects of hardware noise at the effective two-level system description.
We discuss in more detail each of these noise channels in the rest of this subsection.

Depolarizing noise transforms a $n$-qubit quantum state $\rho$ into a completely mixed state, which can be interpreted as a uniform distribution over all $2^n$ computational basis states. The parameter $\lambda_{d}$ indicates the probability in which the output state becomes a completely mixed state~\cite{Nielsen2010}. Its $n$-qubit representation is given by 
\begin{equation}
    \mathcal{E}^{\text{depol}(n)}_{\lambda_d}(\rho) = (1 - \lambda_d) \rho + \frac{\lambda_d \mathbb{I}}{2^n}, 
\end{equation}
where $2^n$ is the dimension of the quantum system and $I/2^n$ is the maximally mixed state. 

Amplitude damping noise stems from the decay of a quantum state $\rho$ from an excited state to the ground state. The parameter $\lambda_a$ is associated with the probability that a state $\ket{\psi}$ decays to the state $\ket{0}$. Higher $\lambda_a$ values correspond to shorter times for this decay. The amplitude damping noise is usually written as a quantum channel 
\begin{equation}
     \mathcal{E}_{\lambda_a}(\rho) = K_1 \rho K_1^\dagger + K_2 \rho K_2^\dagger, 
\end{equation}
where
\begin{equation}
    K_1 = \begin{pmatrix} 1 & 0 \\ 0 & \sqrt{1-\lambda_a} \end{pmatrix}\quad\text{and}\quad
    K_2 = \begin{pmatrix} 0 & \sqrt{\lambda_a} \\ 0 & 0 \end{pmatrix},
\end{equation}
are the Kraus operators~\cite{chuang1997amplitudedamping}. In hardware calibration, amplitude damping is closely associated with the T1 time.

Dephasing noise results from the decoherence of a quantum state, thus reducing the relative phases between the quantum states over time. Similar to amplitude damping described earlier, increasing probability parameter $\lambda_f$ is akin to shortening the time for decoherence to occur. The dephasing noise is written as a quantum channel similar to that for amplitude damping
\begin{equation}
     \mathcal{E}_{\lambda_f}(\rho) = K_1 \rho K_1^\dagger + K_2 \rho K_2^\dagger, 
\end{equation}
where
\begin{equation}
    K_1 = \begin{pmatrix} 1 & 0 \\ 0 & \sqrt{1-\lambda_f} \end{pmatrix}\quad\text{and}\quad
    K_2 = \begin{pmatrix} 0 & 0 \\ 0 & \sqrt{\lambda_f} \end{pmatrix},
\end{equation}
are the associated Kraus operators. Dephasing noise is commonly associated with the T2 time.

The last type of noise that we considered is the readout noise. It represents errors that commonly occur during the measurements of quantum states during quantum computation. Noisy measurement outcomes can be simulated by applying a row-stochastic transition matrix $P$ containing conditional probabilities to the original measurement outcomes. The matrix $P$ is given as
\begin{equation}
    P = \begin{pmatrix} p(0|0) & p(1|0) \\ p(0|1) & p(1|1) \end{pmatrix},
\end{equation}
where the elements $p(y|x)$ represent the probability of the original measurement outcome $x$ being erroneously transformed to outcome $y$. For simplicity, we let the elements $p(0|0) = p(1|1) = 1-\lambda_r$ and $p(1|0) = p(0|1) = \lambda_r$, where $\lambda_r$ represents the average of the readout error rates between $p(0|1)$ and $p(1|0)$.
The readout error channel is essentially a bit flip quantum channel with error-free measurement and can be represented with the following quantum channel
\begin{equation}
    \mathcal{E}_{\lambda_r}(\rho) = K_1 \rho K_1^\dagger + K_2 \rho K_2^\dagger,
\end{equation}
where
\begin{equation}
    K_1 = \begin{pmatrix} \sqrt{1-\lambda_r} & 0 \\ 0 & \sqrt{1-\lambda_r} \end{pmatrix}\quad\text{and}\quad
    K_2 = \begin{pmatrix} 0 & \sqrt{\lambda_r} \\ \sqrt{\lambda_r} & 0 \end{pmatrix},
\end{equation}
are the associated Kraus operators.

We use the quantum computing software Qibo~\cite{efthymiou2021qibo} as a platform to execute quantum circuits.

While additional effects such as coherent errors and crosstalk are not explicitly included in the present noise model, the supervised learning framework allows for the partial absorption of their effective signatures into the fitted parameters of the chosen noise channels. Non-Markovian dynamics, which involve temporal correlations, are not explicitly captured in our approach. However, in the setting considered, these effects do not appear to be dominant, as evidenced by the strong agreement between the emulator and hardware results in Sec.~\ref{sec:results}.

\subsection{Acquire gate-based data via gate set tomography}

The second component of our protocol is a set of gate-based circuits that, when executed on quantum hardware, acquire characterization data which contain footprints of the device noise.
Gate set tomography (GST) \cite{greenbaum2015GST, blume2017GST, nielsen2021GST} is a technique used to characterize gates in the presence of state preparation and measurement errors which can subsequently be used in applications like quantum error mitigation~\cite{endo2018practical}. 
We hypothesize that applying GST on a pre-selected set of native gates would provide the required characterization data for training a neural network which can then be used to extract noise parameters from a real device's GST data.

To carry out GST, one needs to first prepare the states $\rho_k \in \{\ket{0}\bra{0}, \ket{1}\bra{1}, \ket{+}\bra{+}, \ket{y+}\bra{y+} \}^{\otimes n}$, where $n$ is the number of qubits in the system. Subsequently, an $n$-qubit gate may or may not be applied after the prepared state. Expectation values of the Pauli operators are then computed, $M_j = \{\langle I \rangle, \langle X \rangle, \langle Y \rangle, \langle Z \rangle\}^{\otimes n}$.
In the case where no gate is applied after the prepared state, the resulting matrix $g$, often regarded as a calibration matrix, has the following elements
\begin{equation}
    g_{jk} = \text{Tr}(M_j \rho_k).
\end{equation}
When a gate $U$ is applied after the prepared state, the resulting matrix $\mathcal{U}$ contains the following elements
\begin{equation}
    \mathcal{U}_{jk} = \text{Tr}(M_j U \rho_k).
\end{equation}
For single-qubit GST ($n=1$), the resultant $g$ and $\mathcal{U}$ are 4 $\times$ 4 matrices; for two-qubit GST ($n=2$), the resultant $g$ and $\mathcal{U}$ are 16 $\times$ 16 matrices.

Performing GST on quantum hardware requires careful attention to the device's native gates and topology to avoid transpilation-induced circuit modification--such as adding or removing gates--which could compromise the accuracy of GST.
In particular, IQM Garnet uses the Phase Rx gate and CZ gate as its native gates. The Phase Rx gate, hereafter referred to as the PRx gate, can be decomposed into three rotational gates, $\text{PRx}(\theta, \phi) = \text{Rz}(\phi) \text{Rx}(\theta) \text{Rz}(-\phi)$. Together with the CZ gate, they form a universal gate set.
Therefore, to carry out GST on the IQM Garnet device, we modify the GST protocol to only use PRx gates for state preparations and measurements in various bases. This modification specifically allowed GST to be executed with Amazon Braket's verbatim compilation function, bypassing the need for on-device transpilation/compilation to native gates before execution. This enabled us to have full control of target qubits for single-qubit GST as well as control and target qubits for two-qubit GST.

While searching for a suitable gate set for GST, we observed that different types of noise on different gates impact the $g$ and $\mathcal{U}$ matrices with some more similar than others. Amplitude damping noise causes all elements of $g$ and $\mathcal{U}$ to tend towards 1. Dephasing and depolarizing noise may have similar effects on $g$ and $\mathcal{U}$, depending on the choice of gate. Readout noise at a certain parameter resembles maximum depolarizing or dephasing noise as well. We provide a heuristic elaboration in Fig.~\ref{fig.GST_noise_models} in Appendix~\ref{sec.appendix_GST}. The single-qubit PRx gates that form our gate set are listed in Table~\ref{table.PRX_gate_set}.

\begin{table}[h]
\centering
{\renewcommand{\arraystretch}{1.3}
\begin{tabular}{c|cccccccccccccc}
\hline
\textbf{Index} & \textbf{1} & \textbf{2} & \textbf{3} & \textbf{4} & \textbf{5} & \textbf{6} & \textbf{7} & \textbf{8}   & \textbf{9} & \textbf{10} & \textbf{11} & \textbf{12} & \textbf{13} & \textbf{14} \\ \hline
$\theta$       & $\pi$ & $\frac{\pi}{2}$  & $\frac{\pi}{2}$ & $\frac{\pi}{3}$ & $\frac{\pi}{3}$ & $\frac{\pi}{3}$ & $\frac{\pi}{4}$ & $\frac{\pi}{4}$ & 0 & $-\frac{\pi}{2}$ & $\frac{\pi}{4}$ & $\frac{\pi}{7}$ & $\frac{\pi}{7}$ & $\frac{\pi}{7}$ \\
$\phi$         & 0     & $-\frac{\pi}{2}$ & $\frac{\pi}{2}$ & $\frac{\pi}{3}$ & $\frac{\pi}{4}$ & $\frac{\pi}{7}$ & $\frac{\pi}{3}$ & $\frac{\pi}{4}$ & 0 & 0                & $\frac{\pi}{7}$ & $\frac{\pi}{3}$ & $\frac{\pi}{4}$ & $\frac{\pi}{7}$ \\ \hline
\end{tabular} 
}
\caption{List of PRx gate angles for 1 qubit gate set.}
\label{table.PRX_gate_set}
\end{table}

The two-qubit gate set only consists of the CZ gate as it is the only native two-qubit gate of IQM Garnet. Further information about the IQM Garnet device, such as the topology, calibration data details, and costs, can be found in Appendix~\ref{sec.appendix_garnet}.

We emphasize that the GST procedures employed in this work are restricted to single-qubit and two-qubit gate sets, applied locally across qubits and their couplings. As such, the overall scaling of the protocol is governed by the number of qubits and the connectivity of the device (i.e., the number of coupled qubits), rather than the exponential scaling associated with full n-qubit GST. The total experimental cost is therefore determined by the number of local GST experiments and the measurement shots required for each instance.

\subsection{Train neural networks to predict noise parameters}
\label{sec.training_and_evaluating}

The third and final component of our protocol is a machine learning framework that can be used to predict noise parameters from GST data. 
For simplicity and ease of implementation, we chose two feed-forward neural networks: one to be trained on single-qubit GST data (NN-1Q) and the other to be trained on two-qubit GST data (NN-2Q).
The details of the architecture of NN-1Q and NN-2Q can be found in Table~\ref{table.NN-1Q_NN-2Q_params}. In addition, all data used to train NN-1Q as well as the generated data for NN-2Q, the weights of NN-1Q and NN-2Q, and the standard scalers applied to both neural networks are available at Ref.~\cite{GTH_HKMC_2025}.

The evaluation process occurs in two stages. The first stage happens in real-time during the training of the neural network. Using a validation split parameter that determines the proportion of samples allocated to compute the validation loss function, we monitor the validation loss alongside the training loss. This feedback provides real-time analysis of overfitting (when validation loss increases while training loss decreases) or underfitting (when both validation loss and training loss remain high) of the model.

The second stage takes place after training is complete. Here, the performance of a neural network is evaluated by comparing the neural network's predictions against ground truth values. This is done by performing GST again using predicted values of $\Lambda_\text{true}$ or $\zeta_\text{true}$ in two settings: (a) using predetermined values of $\Lambda_\text{true}$ or $\Lambda_\text{true}$ as those used during training and (b) using predefined values of $\Lambda_\text{true}$ and $\zeta_\text{true}$ with finer granularity than those used in training.
In both cases, the newly generated GST data is fed into the neural network and the proximity of the predicted values ($\Lambda_\text{predicted}$ to $\Lambda_\text{true}$ and $\zeta_\text{predicted}$ to $\zeta_\text{true}$) is analyzed. The mean squared error (MSE) quantifies the distance between ground truth and predicted values. In addition, a ``predicted vs true" plot provides visual confirmation.

\begin{table}[]
\centering
\begin{tabular}{l|l|l}
\hline
\textbf{Parameters} & \textbf{NN-1Q} & \textbf{NN-2Q} \\ \hline
\begin{tabular}[c]{@{}l@{}}No. of intermediate\\ layers (excluding\\ output)\end{tabular} & 2                                                                     & 2                                                                     \\ \hline
\begin{tabular}[c]{@{}l@{}}No. of nodes in\\ intermediate layer\end{tabular}              & 128                                                                   & 64                                                                    \\ \hline
\begin{tabular}[c]{@{}l@{}}No. of nodes in\\ output layer\end{tabular}                    & 4                                                                     & 1                                                                     \\ \hline
\begin{tabular}[c]{@{}l@{}}Activation function\\ for intermediate\\ layers\end{tabular}   & ReLU                                                                  & ReLU                                                                  \\ \hline
\begin{tabular}[c]{@{}l@{}}Activation function\\ for output layer\end{tabular}            & \begin{tabular}[c]{@{}l@{}}Custom sigmoid\\ ($\alpha=8$)\end{tabular} & \begin{tabular}[c]{@{}l@{}}Custom sigmoid\\ ($\alpha=8$)\end{tabular} \\ \hline
Optimizer                                                                                 & Adam                                                                  & Adam                                                                  \\ \hline
Learning rate                                                                             & $10^{-4}$                                                                  & $10^{-4}$                                                                  \\ \hline
Loss function                                                                             & \begin{tabular}[c]{@{}l@{}}Mean squared\\ error\end{tabular}          & \begin{tabular}[c]{@{}l@{}}Mean squared\\ error\end{tabular}          \\ \hline
L2 regularization                                                                         & $5 \times 10^{-5}$                                                                  & $5 \times 10^{-6}$                                                                  \\ \hline
Dropout                                                                                   & $5 \times 10^{-5}$                                                                  & $5 \times 10^{-6}$                                                                  \\ \hline
Batch size                                                                                & 64                                                                    & 32                                                                    \\ \hline
Validation split                                                                          & 0.2                                                                   & 0.2                                                                   \\ \hline
\begin{tabular}[c]{@{}l@{}}Recommended  no. of\\ training episodes\end{tabular}           & 100                                                                   & 2,500  \\ \hline                                                              
\end{tabular}
\caption{Parameters for NN-1Q and NN-2Q.}
\label{table.NN-1Q_NN-2Q_params}
\end{table}

\subsection{Recipe for constructing $\mathcal{N}^{(2)}_{\Lambda_{q_i, q_j}}$}
\label{sec.recipe}

We outline these steps in accordance with Fig.~\ref{fig.workflow} to obtain the GTH noise model for qubits $q_i$ and $q_j$.
\begin{enumerate}
    \item Define $\Lambda = \{\lambda_d, \lambda_a, \lambda_f, \lambda_r \}$, representing single-qubit depolarizing noise, amplitude damping noise, dephasing noise, and readout noise parameters, respectively. 
    \item Define a discrete range for $\Lambda$.
    \item For each value in the range for $\Lambda$, use single-qubit noise model 
    $\mathcal{N}^{(1)}_{\Lambda}$ (Eq.~\ref{eq.noise_model_1qb}) to generate single-qubit simulated GST data. The full dataset is then used to train NN-1Q.
    \item Perform single-qubit GST for qubit $q_i$ on hardware, input the results into the trained NN-1Q to obtain predicted noise $\Lambda_{q_i}$.
    \item Repeat Step 4 on qubit $q_j$ (which is coupled to $q_i$) to get $\Lambda_{q_j}$.
    \item Construct two-qubit composite noise model $\mathcal{N}^{(2)}_{\Lambda_{q_i, q_j}}$ (Eq.~\ref{eq.noise_model_2qb}) using $\Lambda_{q_i}$, $\Lambda_{q_j}$, and $\zeta$. Recall that $\zeta$ is an arbitrary value for the two-qubit depolarizing noise parameter.
    \item Define a discrete range for $\zeta$.
    \item For each value in the range for $\zeta$, use $\mathcal{N}^{(1)}_{\Lambda=\Lambda_{q_i}} \circ \mathcal{N}^{(1)}_{\Lambda=\Lambda_{q_j}} \circ \mathcal{E}^{\text{depol}(2)}_{\zeta}$ to generate two-qubit simulated GST data. The full dataset is used to train NN-2Q.
    \item Perform two-qubit GST for qubit $q_i$ and $q_j$ on hardware, input the results into the trained NN-2Q to obtain the predicted two-qubit depolarizing noise $\zeta_{q_i, q_j}$.
    \item Finally, form the GTH noise model $\mathcal{N}^{(2)}_{\Lambda_{q_i, q_j}} = \mathcal{N}^{(1)}_{\Lambda=\Lambda_{q_i}} \circ \mathcal{N}^{(1)}_{\Lambda=\Lambda_{q_j}} \circ \mathcal{E}^{\rm depol(2)}_{\zeta=\zeta_{q_i, q_j}}$. 
\end{enumerate}

We provide a more detailed explanation of the training procedure, including the choice of step sizes for $\Lambda$ and $\zeta$, in Appendix~\ref{sec.appendix_recipe}.

\section{Results}~\label{sec:results}

In this section, we first present the performance of NN-1Q and NN-2Q before looking at how $\mathcal{N}^{(2)}_{\Lambda_{q_i, q_j}}$ is benchmarked against hardware results with a quantum chemistry example.

Our work was limited by the amount of credits, with enough budget to explore a two-qubit system thoroughly. Hence, we choose qubits 1 and 4 of the IQM Garnet device to run our two-qubit system on as well as characterize. More information about the calculation of the costs can be found in Appendix.~\ref{sec.appendix_GST_time_costs}.

\subsection{Performance of NN-1Q and NN-2Q}

\begin{figure}[t]
    \includegraphics[width=1.0\linewidth]{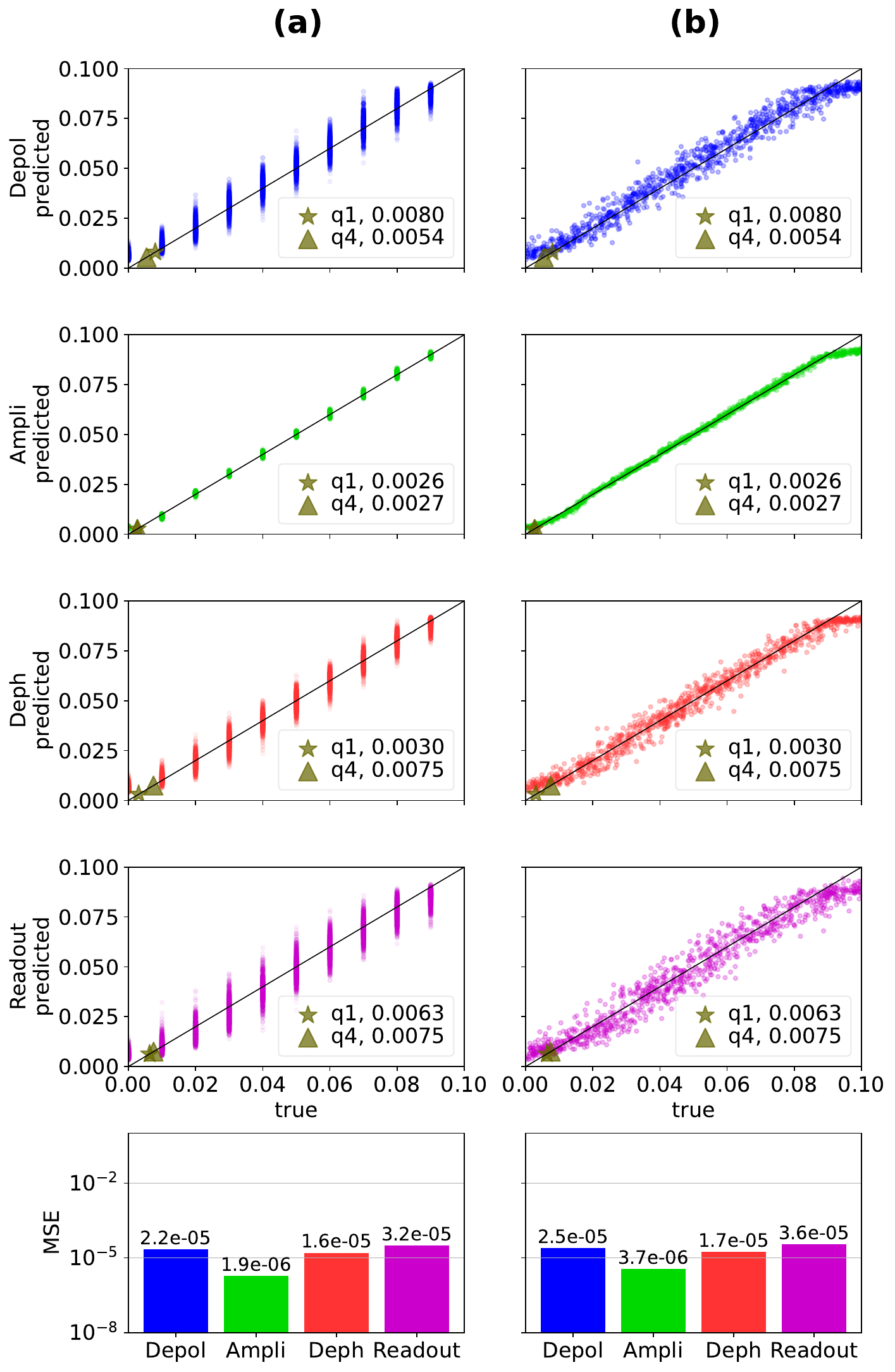}
    \caption{Evaluation of NN-1Q using (a) unseen data with predetermined ground truths values as those used during training and (b) predefined ground truths with finer granularity than those used in training. The predicted values from the IQM GST data are plotted on the $y=x$ line for visual reference. The bottom bar charts display the mean squared error (MSE) of the both evaluation settings.}
    \label{fig.NN-1Q_results}  
\end{figure}

\begin{figure}[t]
    \includegraphics[width=1.0\linewidth]{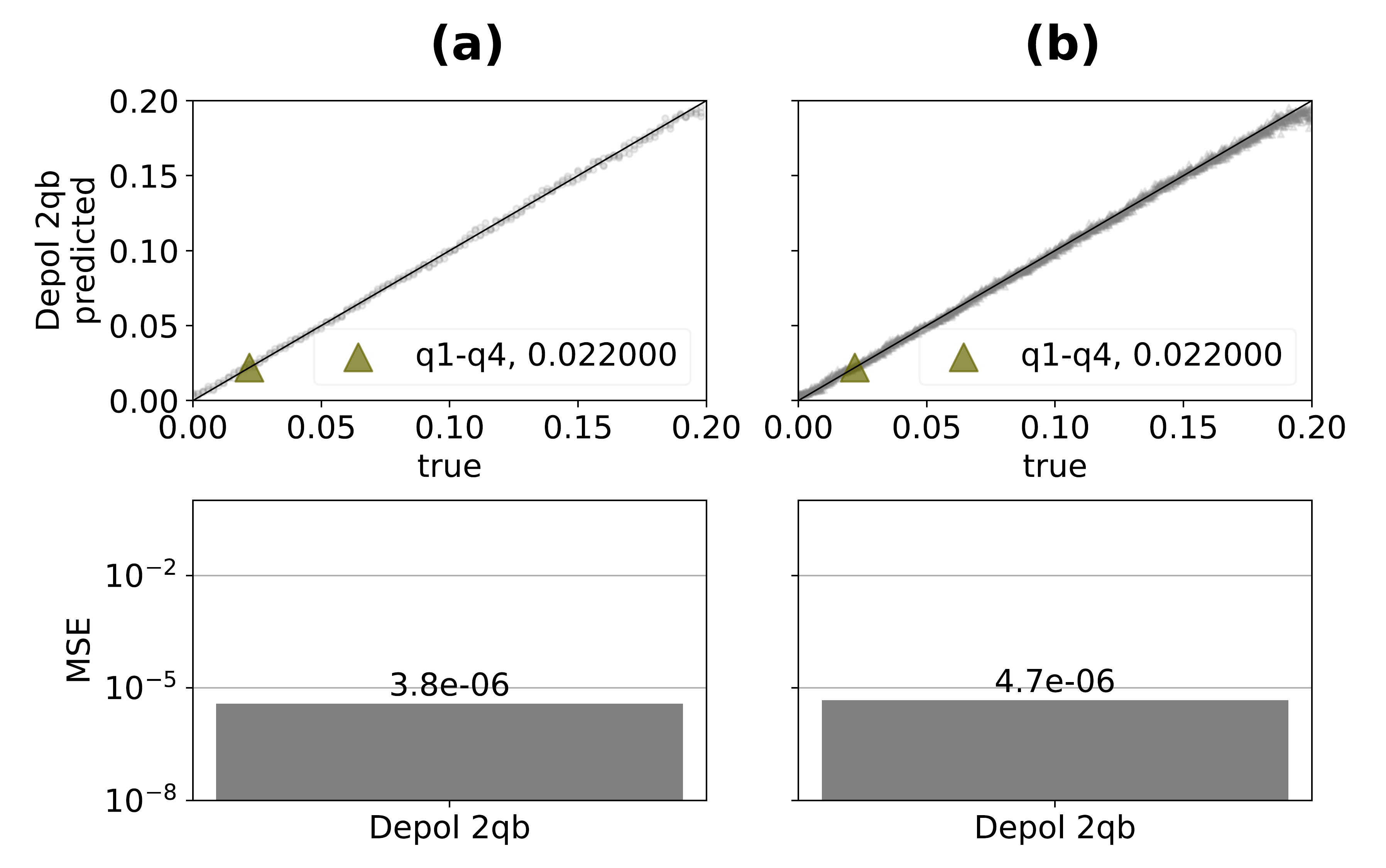}
    \caption{Evaluation of NN-2Q using (a) unseen data with predetermined ground truths values as those used during training and (b) predefined ground truths with finer granularity than those used in training. The predicted values from the IQM GST data are plotted on the $y=x$ line for visual reference but are close to zero as seen in the legends. The bottom bar charts display the mean squared error (MSE) of the both evaluation settings.}
    \label{fig.NN-2Q_results}
\end{figure}

\begin{table}[]
\centering
\begin{tabular}{c|c|c}
\hline
\hspace{0.5cm} \textbf{Single-qubit} \hspace{0.5cm} & \textbf{$q_1$} & \textbf{$q_4$} \\ \hline
$\lambda_d$ & \hspace{0.5cm} 0.00799 \hspace{0.5cm} & \hspace{0.5cm} 0.00541 \hspace{0.5cm} \\
$\lambda_a$ & 0.00260 & 0.00270 \\
$\lambda_f$ & 0.00305 & 0.00747 \\
$\lambda_r$ & 0.00632 & 0.00749 \\ \hline
\textbf{Two-qubit} & \multicolumn{2}{c}{\textbf{$q_1$-$q_4$}} \\ \hline
$\zeta$ & \multicolumn{2}{c}{0.0220} \\ \hline
\end{tabular}
\caption{Predictions for IQM Garnet's qubit $q_1$ and qubit $q_4$ individually by NN-1Q and coupled qubits $q_1$ and $q_4$ by NN-2Q to 5 decimal places.}
\label{table.NN-1Q_NN-2Q_predicted_IQM}
\end{table}

Following the protocol in Fig.~\ref{fig.workflow}, we performed single-qubit GST on IQM Garnet with the 4 gates listed in Table~\ref{table.PRX_gate_set} on qubits $q_1$ and $q_4$ individually followed by two-qubit GST, also on qubits $q_1$ and $q_4$. The calibration data for the days we performed GST was concurrently obtained and stored.

The single-qubit GST was performed with 10,000 shots while the two-qubit GST was performed with 1,000 shots. A larger shot count was necessary to reduce statistical fluctuations in the GST data, which could otherwise provide misleading higher error-ridden input data into neural network, affecting the predictions. Despite this, we found that using 1,000 shots for two-qubit GST was sufficient to predict the two-qubit depolarizing noise.

We plot the evaluation of NN-1Q in Fig.~\ref{fig.NN-1Q_results} and NN-2Q in Fig.~\ref{fig.NN-2Q_results}, illustrating the ability of NN-1Q and NN-2Q in the two types of evaluation described in Section~\ref{sec.training_and_evaluating}. In the ``predicted versus true" plots for NN-1Q, we notice that the network's predictions are scattered along the diagonal. This can be attributed to the impact of the finite shots used (10,000 shots) to generate the GST training data.

The prediction accuracy is quantified using the MSE, with the corresponding plots shown in the bar charts. The largest mean squared error is in the order $O(10^{-5})$, which is sufficiently small to provide a reasonable level of confidence.

Visually, we note that NN-2Q performs considerably better than the four single-qubit noise models as the predicted values have less scattering along the diagonal. It is quantitatively confirmed by a low mean squared error in the order of $O(10^{-6})$. This is solely due to NN-2Q being used to predict only one type of noise--the two-qubit depolarizing noise model. 

A summary of the predictions for $\Lambda_{q_1}$, $\Lambda_{q_1}$, and $\zeta_{q_1, q_4}$ is listed in Table.~\ref{table.NN-1Q_NN-2Q_predicted_IQM}. These predictions for the noise models are used as inputs for in the composite noise model $\mathcal{N}^{(2)}_{\Lambda_{q_i, q_j}}$.

\begin{figure*}
    \parbox{\linewidth}{\raggedright
        \textbf{(a)}\\
        \includegraphics[width=0.95\linewidth]{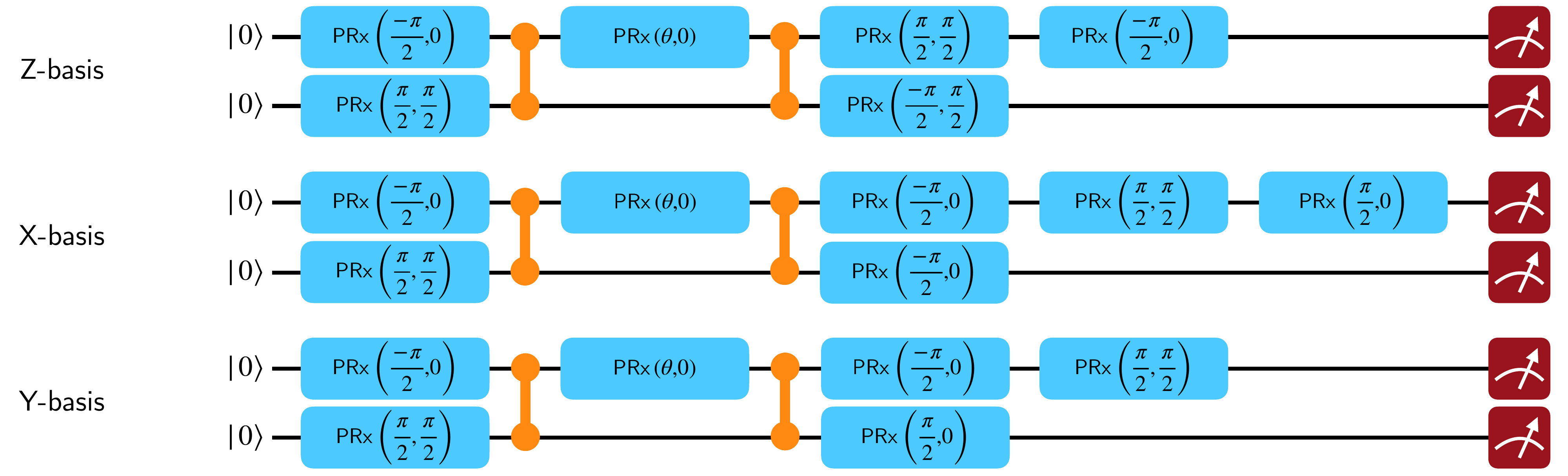} \\
        \textbf{(b)}\\
        \includegraphics[width=1\linewidth]{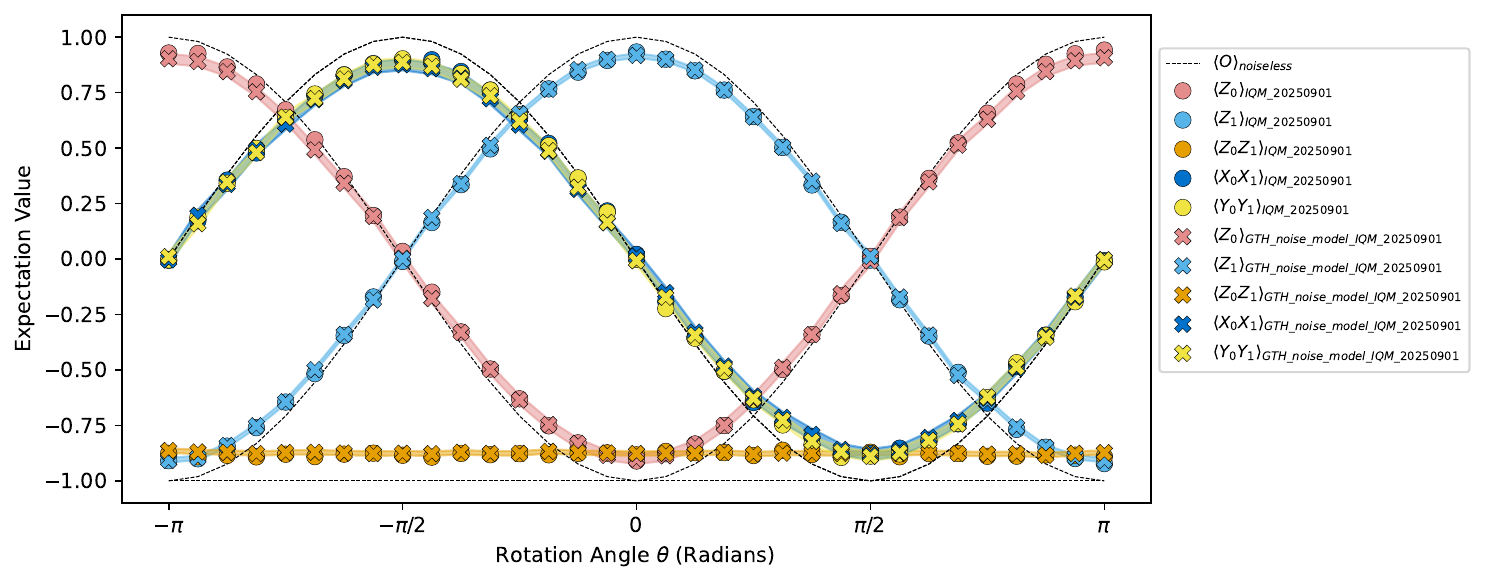} \\
    }
    \caption{
    (a) Transpiled UCC circuits to IQM Garnet native gates for measurement in the $Z$, $Y$, and $X$ bases. (b) Adjusting the parameter $\theta$ from $-\pi$ to $\pi$ and executing them on IQM Garnet as well as with our noise model, we obtain the corresponding expectation values for the Pauli terms. Shaded regions are used in place of error bars to reduce visual clutter. They represent the root mean square of the absolute difference between expectation values obtained from IQM Garnet and the GTH noise model across all rotation angles. The mean absolute difference between composite noise model values and device values is $0.01385$. Note that $\langle Y_0 Y_1 \rangle$ and $\langle X_0 X_1 \rangle$ are almost identical for noiseless, IQM, and GTH noise model, resulting in a large degree of overlap among the scatter points.
    The above UCC sweep was executed on the IQM Garnet on 2025-09-01, on the same day as and right after executing the GST circuits required for obtaining the GTH noise parameters to minimize device noise drift.
    }
    \label{fig.UCC_circuits}
\end{figure*}

\subsection{Application of the GTH noise model $\mathcal{N}^{(2)}_{\Lambda_{q_i, q_j}}$: Quantum Chemistry}

\subsubsection{Unitary Coupled Cluster energy of the H$_2$ molecule}
\label{sec.UCC}

We present here a test of $\mathcal{N}^{(2)}_{\Lambda_{q_i, q_j}}$ on an example quantum chemistry calculation, namely the unitary coupled cluster (UCC) energy of the H$_2$ molecule with atomic separation of 0.70\AA \ using the STO-6G basis set, as demonstrated by O’Malley et al.~\cite{omalley2016h2}. We provide a detailed description of the UCC method in Appendix \ref{sec.appendix_ucc}, and shall only provide the important experimental details here. The second quantized electronic Hamiltonian for the above system is transformed to the following two-qubit effective Hamiltonian after Bravyi-Kitaev~\cite{bravyi2002bk} mapping:
\begin{equation} \label{eq.H2_ham}
  \begin{split}
    H_\text{eff} = &-0.4584\mathbb{I} + 0.3593Z_0 - 0.4826Z_1 + 0.5818Z_0Z_1 \\ 
    &+0.0896X_0X_1 +0.0896Y_0Y_1.
  \end{split}
\end{equation}

The ground state for the above Hamiltonian as predicted by UCC theory is
\begin{equation}
    |\psi(\theta)\rangle = e^{-i\theta X_0 Y_1}|\phi\rangle.
\end{equation}
The state $|\phi\rangle$ is a reference state for the electrons in the molecular system. We adopt the Hartree-Fock state as $|\phi\rangle$, which is encoded big-endian here as $|10\rangle$; the first orbital is occupied and second orbital is vacant, or virtual. Using a parameterized ansatz $|\psi(\theta)\rangle = U(\theta)|\phi\rangle$ with parameter vector $\theta$ on the quantum circuit used by O'Malley et al.~\cite{omalley2016h2} the corresponding circuits in the basis to be measured ($Z$, $X$, or $Y$), transpiled to IQM Garnet's native gates, is given in Fig.~\ref{fig.UCC_circuits}(a).

\begin{table}
\centering
\begin{tabular}{c|c|c}
\hline
 & $\Delta E_0$ & $\Delta \text{sweep}$ \\ \hline
\hspace{0.5cm} GTH noise model \hspace{0.5cm} & \hspace{0.5cm} 0.128\% \hspace{0.5cm} & \hspace{0.5cm} 0.01385 \hspace{0.5cm} \\
IBMQNoiseModel & 4.336\% & 0.01961 \\ \hline
\end{tabular}
\caption{Two evaluation metrics $\Delta E_0$ and $\Delta \text{sweep}$ are given for the GTH noise model and the IBMQNoiseModel, which requires device calibration data such as T1 and T2 times. Firstly, $\Delta E_0$ denotes the percentage relative error of the ground state energy with respect to IQM Garnet ($E_0 = -0.9993276$) evaluated at $\theta=\theta^*$ (GTH noise model: $E_0 = -1.0006050$; IBMQNoiseModel: $E_0 = -1.0426578$). Secondly, $\Delta \text{sweep}$ represents the mean absolute difference of the UCC sweep over the range $-\pi \leq \theta \leq \pi$ relative to IQM Garnet.}
\label{table.UCC_results}
\end{table}

Adopting a variational approach using state-vector simulation (noiseless simulation) to minimize the UCC energy for H$_2$,
\begin{equation}
    E_0 = \min_{\theta} \frac{\langle \psi(\theta) |H_\text{eff}|\psi(\theta)\rangle} {\langle \psi(\theta) | \psi(\theta) \rangle},
\end{equation}
the optimal parameter was found to be $\theta^* = 0.2097$. 
Substituting $\theta^*$ into the three circuits given in Fig.~\ref{fig.UCC_circuits}(a), we then ran these circuits using IQM Garnet with 10,000 shots and computed the expectation value. 
Separately, these exact circuits were then subjected to the noise model $\mathcal{N}^{(2)}_{\Lambda_{q_1, q_4}} = \mathcal{N}^{(1)}_{\Lambda=\Lambda_{q_1}} \circ \mathcal{N}^{(1)}_{\Lambda=\Lambda_{q_4}} \circ \mathcal{E}_{\zeta=\zeta_{q_1, q_4}}^{\rm depol(2)}$ and simulated classically.
The results are given in Table~\ref{table.UCC_results}. 
The percentage relative error in $E_0$ between the GTH noise model and IQM Garnet was found to be a mere 0.128\%, highlighting the accuracy of the GTH noise model in emulating the behavior of the IQM Garnet. Here, the percentage relative error is calculated using the expression $\tfrac{|B-A|}{|A|} \times 100\%$ with $A$ representing IQM Garnet and $B$ representing a noise model (either the GTH or the IBMQNoiseModel).

To evaluate the relevance of our GTH noise model, we performed the simulations with the IBMQNoiseModel~\cite{IBMQNoiseModel}, which is a noise model that is based upon thermal relaxation, depolarizing, and readout noise channels, and allows direct incorporation of calibration parameters (e.g., T1, T2, readout errors, and depolarizing noise) from the IQM Garnet device. We emphasize that the IBMQNoiseModel is used here solely as a reference baseline based on device characterization and is not intended as a universal or hardware-agnostic standard for noise modeling. The calibration data was obtained on the same day when the 1-qubit GST, 2-qubit GST, and UCC sweep were successively evaluated on the IQM Garnet device. The percentage relative error in $E_0$ between the IBMQNoiseModel and IQM Garnet was found to be 4.336\%, whereas the percentage relative error in $E_0$ between our GTH noise model and IQM Garnet was 0.128\%. This indicates that the GTH noise model has a higher accuracy in replicating IQM Garnet’s noise.

We emphasize here that calculating the UCC energy $E_0$ of the H$_2$ molecule involves computing five distinct expectation values, $Z_0, Z_1, Z_0 Z_1, X_0 X_1, Y_0 Y_1$, from three separate circuits (shown in Fig.~\ref{fig.UCC_circuits}(a)). Therefore, it is unlikely that the striking agreement of the GTH noise model with the actual hardware is merely due to coincidence.

For a more thorough assessment, we ran all three circuits in Fig.~\ref{fig.UCC_circuits}(a), 
sweeping $\theta$ across a range of values from $-\pi$ to $\pi$ in 33 equal steps on both IQM Garnet as well as our GTH noise model emulator.
To minimize the noise drift of the QPU, the sweep were conducted on the IQM Garnet immediately after executing the 1- and 2-qubit GST.
The sweep results are post-processed to compute the expectation values for each Pauli term in $H_\text{eff}$ (Eq.~\ref{eq.H2_ham}) and illustrated in Fig.~\ref{fig.UCC_circuits}(b), where shaded regions (instead of error bars) indicate the root mean square of the absolute difference between IQM Garnet's results and the GTH noise model's results for the respective Pauli terms across all rotation angles.
Our GTH noise model is seen to emulate the IQM Garnet device very well, with a mean absolute difference of 0.01385 for the expectation value across all Pauli terms in the Hamiltonian.
Similarly, the IBMQNoiseModel achieves a mean absolute difference of 0.01961. 
As shown also in Table~\ref{table.UCC_results}, our GTH noise model outperforms the 
IBMQNoiseModel with up-to-date calibration parameters in terms of emulating the IQM Garnet device, achieving significantly lower values for both the UCC relative energy difference as well as the UCC sweep mean absolute difference (see Appendix~\ref{sec.appendix_gth_vs_ibmqnoisemodel} for additional details).

These remarkable results confirm that
ML-based noise models trained on GST data are capable of emulating device noise, thereby validating our protocol's feasibility to construct faithful NISQ device emulators.

We observe that the accuracy of the GTH noise model degrades when there is a delay between single-qubit and two-qubit GST data acquisition and the execution of the UCC circuits, which we attribute to temporal drift in the device noise characteristics. This indicates that the learned noise model is effectively time-local, capturing the device behavior at the moment of characterization. When GST and UCC circuits are performed on different days, such drift becomes more pronounced. To minimize this effect and maintain high-fidelity emulation, the single-qubit GST, two-qubit GST, and application circuits should be executed as closely in time as possible, ideally within the same day.

\section{Conclusion}\label{sec:conclusions}

In this work, we constructed the GTH noise model and demonstrated its potential to accurately capture the noise on the IQM Garnet device. Our emulation of the UCC circuits for H$_2$ with the GTH noise model at the optimal angle $\theta^*$ differs from IQM Garnet results by a mere 0.128\% in relative difference. Sweeping through angles $-\pi \leq \theta \leq \pi$, the GTH noise model gave results with a mean absolute difference of 0.01385 from the actual hardware run on IQM Garnet.
Through achieving high-fidelity emulation of an actual quantum processor, we use the GTH noise model as a specific realization to demonstrate how general users can apply our proposed ML-driven, generalized gate-based protocol to construct a heuristic but yet effective noise model for any quantum computer of their choice.

The effectiveness of the GTH noise model shows that not only does the data obtained from GST leave substantial noise footprints of the hardware, we demonstrate how general users can exploit machine learning to turn the data used for characterizing the performance of quantum hardware to build a noise model that emulates the hardware itself. Additionally, our work shows that to closely approximate actual hardware noise, input from an underlying physical model of the device is not required and it is in fact possible to do so with heuristic noise models that are sufficiently complex. Unlike such noise models that are based upon device calibration parameters, the noise parameters of our GTH noise model are essentially fitting parameters and do not necessarily correspond to any physical characteristics of the hardware. The success of this preliminary work opens the door 
for a complementary approach for realistic qubit characterization and device-noise modeling that may be just as or even more accurate in reproducing hardware results.
Substantial improvements can be made by utilizing other candidate heuristic noise models, further optimizing the neural network design, as well as designing other gate-based device characterization circuits beyond GST. 

Despite limits in credit budget, our work is paramount to laying the foundation for general users to gain access to accurate quantum device emulation.
The protocol can be extended in a straightforward manner from a two-qubit case to an $n$-qubit case $(n>2)$ by applying single-qubit and two-qubit GST across single qubits and their couplings, following the qubit connectivity of the quantum device, with scaling determined solely by the number of single and two qubit GSTs rather than exponential scaling associated with global $n$-qubit GST. This entails allocating sufficient budget to perform GST on additional qubits (preferably coupled qubits), collecting the GST data, and using the neural networks to predict the noise parameters. This opens the possibility for exploration of multi-qubit systems. While it is possible to emulate the noise characteristics for a larger portion of the NISQ device, we recognize that the GST experiments required in our current protocol will nonetheless incur substantial costs to general users.
On the other hand, users with limited quantum compute time such as those under the IBMQ `Open Plan' will be unable to execute our protocol due to the number of quantum circuits that need to be executed to collect GST data (see Appendix~\ref{sec.appendix_GST_time_costs} for detailed discussion).

Future work will explore improvements to reduce such costs, potentially through incorporating less costly experiments such as randomized benchmarking. 
This will enable general users to build NISQ device emulators with more qubits, allowing for simulations of larger and more complex quantum algorithms that can be meaningfully compared to hardware results.
We will also explore more comprehensive benchmarks to quantify emulator performance.
That said, we consider our current example--calculating the UCC energy of the H$_2$ molecule, which requires measurements in all three Pauli bases--to be sufficiently comprehensive and conclusive in this preliminary work.

\section*{Acknowledgements} This research is supported by the National Research Foundation, Singapore and A*STAR under the Quantum Engineering Programme, NRF2021-QEP2-02-P01, NRF2021-QEP2-02-P02, NRF2021-QEP2-02-P03. We thank Amazon Web Services for cloud quantum computing access. This work was also supported by QNIX NQSTI-PNRR CUP H43C22000870001 (S.C.).

\bibliographystyle{ieeetr}
\bibliography{biblio}

\newpage

\appendix
\begin{widetext}

\section{Unitary Coupled Cluster}
\label{sec.appendix_ucc}

For a molecular system of clamped nuclei (Born-Oppenheimer approximation), the electronic Hamiltonian in second quantization using creation and annihilation operators $a^\dagger$ and $a$ is written as
\begin{equation}
\label{eq:electronichamiltonian}
    H_{\mathrm{elec}} = \sum_{pq}h_{pq}a^\dagger_pa_q + \frac12\sum_{pqrs}(pr|qs)a^\dagger_pa^\dagger_q a_s a_r,
\end{equation}
where $h_{pq}$ and $(pr|qs)$ are one- and two-electron integrals obtained from PySCF, $p,q,r,s$ are general orbital indices. 
The Bravyi-Kitaev transformation of the fermionic Hamiltonian for the H$_2$ molecule in the minimal STO-6G basis is given in Ref.~\cite{omalley2016h2} as
\begin{equation}    
\begin{aligned}
\label{eq.BK_hamiltonian}
    H_\text{BK} = & f_0 \mathbb{I} + f_1 Z_0 + f_2 Z_1 + f_3 Z_2 + f_1 Z_0 Z_1  \\
    & + f_4 Z_0 Z_2 + f_5 Z_1 Z_3 + f_6 \textbf{\textit{X}}_0 Z_1 \textbf{\textit{X}}_2 + f_6 \textbf{\textit{Y}}_0 Z_1 \textbf{\textit{Y}}_2 \\
    & + f_7 Z_0 Z_1 Z_2 + f_4 Z_0 Z_2 Z_3 + f_3 Z_1 Z_2 Z_3 \\
    & + f_6 \textbf{\textit{X}}_0 Z_1 \textbf{\textit{X}}_2 Z_3 + f_6 \textbf{\textit{Y}}_0 Z_1 \textbf{\textit{Y}}_2 Z_3 + f_7 Z_0 Z_1 Z_2 Z_3.
\end{aligned}
\end{equation}

The values $f_i$ are weights that are evaluated from the one- and two-electron integrals for a given geometry (bond length) of the H$_2$ molecule. The authors for that paper observed that ``this Hamiltonian acts off-diagonally on only two qubits, (the ones having tensor factors or 0 and 2), those set in bold (in Eq.~\ref{eq.BK_hamiltonian}." , and that ``the Hamiltonian stabilizes qubits 1 and 3 so that they are never flipped throughout the simulation". The Hamiltonian in Eq.~\ref{eq:electronichamiltonian} is thus reduced using symmetry to the following effective Hamiltonian,
\begin{equation}
    H_{\mathrm{eff}} = g_0 \mathbb{I} + g_1 Z_0 + g_2 Z_1 + g_3 Z_0 Z_1 + g_4 X_0 X_1 + g_5 Y_0 Y_1.
\end{equation}

As a result of the reduction of the Hamiltonian, qubits 0 and 1 are reduced in the new scheme to qubit 0, and qubits 2 and 3 are reduced to qubit 1. The doubles excitation in the Unitary Coupled Cluster ansatz which there is non-zero amplitude, i.e. $a^\dagger_2 a^\dagger_3 a_1 a_0 - a^\dagger_0 a^\dagger_1 a_3 a_2$ is thus reduced to $a^\dagger_1 a_0 - a^\dagger_0 a_1$.

The electronic state of this molecular system is represented using the Unitary Coupled Cluster (UCC) ansatz, a variant of the gold-standard Coupled Cluster theory in quantum chemistry. Cluster generators $T_n(\theta)$ with excitation parameter $\theta$ effect n-tuple excitation of electrons from occupied orbitals $i,j,...$ to unoccupied (virtual) orbitals $a,b...$, and for singles and doubles excitations they are
\begin{align}
    T_1(\theta_{ia}) &= \sum_{ia}\theta_{ia}a^\dagger_a a_i \\
    T_2(\theta_{iajb}) &= \frac14\sum_{iajb}\theta_{iajb}a^\dagger_a a_i a^\dagger_b a_j.
\end{align}

A cluster operator $T$ is generally
\begin{equation}
    T(\theta) = T_1(\theta_{ia}) + T_2(\theta_{iajb}) + \dots + T_m(\theta_{(m)}),
\end{equation}
where $m$ is the highest level of excitations and $(m)$ indicates the excitations for cluster generator $T_m$, e.g. for singles, $m=1$, $(m)\rightarrow ia$.

Starting from a mean-field starting state, namely the Hartree-Fock state $|\phi\rangle$ obtained from a classical computation, the UCC ansatz is implemented as
\begin{align}
    |\psi(\theta)\rangle &= U(\theta)|\phi\rangle \\
    &= e^{(T(\theta)-T^\dagger(\theta))}|\phi\rangle.
\end{align}

The electronic energy is variationally minimized using the state-vector simulation to obtain the UCC energy,
\begin{equation}
    E_0 = \min_{\theta} \frac{\langle \psi(\theta) |H_\mathrm{eff}|\psi(\theta)\rangle} {\langle \psi(\theta) | \psi(\theta) \rangle},
\end{equation}
with optimal parameter found to be $\theta^* = 0.2097$.

In addition, we plot the entanglement entropy (von Neumann entropy) in the same range, as shown in Fig.~\ref{fig.entanglement_entropy_sweep}.
\begin{figure}
    \centering
    \includegraphics[width=1.0\linewidth]{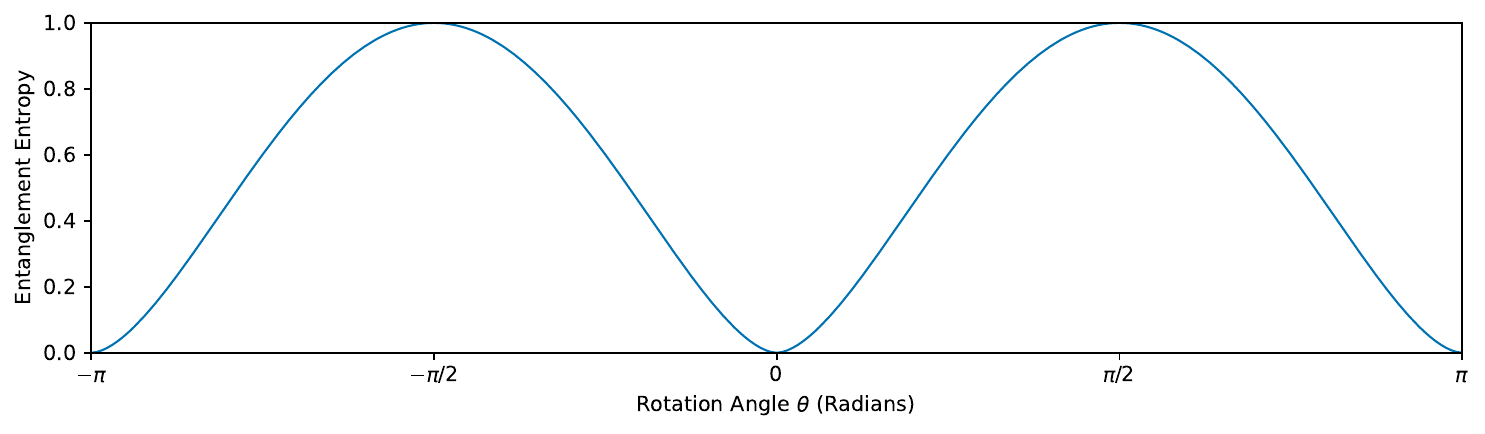}
    \caption{The von Neumann entropy $S(\rho) = -\text{Tr} (\rho \log \rho)$ gives the value of entanglement entropy ($0 \leq S \leq 1$). Here, the $S(\rho)$ illustrates the degree of entanglement while $\theta$ is varied from $-\pi$ to $\pi$. The amount of entanglement entropy at the optimal solution $\theta^* = 0.2097$ is $S(\theta^*) = 0.08705$.}
    \label{fig.entanglement_entropy_sweep}
\end{figure}

The main text presents results from successive back-to-back execution of 1-qubit GST, 2-qubit GST, and UCC sweeps performed on the IQM Garnet within a three-hour window, during which noise drift was minimized as much as possible.
In this section, we provide a complementary comparison using delayed execution, where the 1-qubit GST, 2-qubit GST, and UCC sweeps were carried out on separate days, inevitably introducing additional noise drift.

Table~\ref{table.UCC_side_by_side} shows the relative differences between successive and delayed executions, Fig.~\ref{fig.NN2Q_side_by_side} illustrates the difference in NN-2Q evaluations, and Fig.~\ref{fig.UCC_sweep_side_by_side} shows the plots of UCC sweep for successive and delayed executions.
As we expect, the performance of the GTH noise model is significantly better under successive execution than delayed execution of the circuits, as is indicated by the lower UCC relative energy difference as well as the UCC sweep mean absolute difference. Nonetheless, the GTH noise model retains high accuracy even with delayed circuit execution, suggesting relatively low noise drift and high consistency of the IQM Garnet device.

The mean absolute difference is computed in the following manner. First, the mean absolute error between two pairs of vectors $P = (p_1, p_2, ..., p_n)$ and $Q = (q_1, q_2, ..., q_n)$ is defined as 
\begin{equation}
    \text{MAE}(P, Q) = \frac{1}{n} \sum_{i=1}^n | p_i - q_i |.
\end{equation}
Next, the mean absolute difference is
\begin{align}
    \text{MAD} = \frac{1}{5} \Big( & \text{MAE}(P_{Z_0}, Q_{Z_0}) + \text{MAE}(P_{Z_1}, Q_{Z_1}) + \\
    & \text{MAE}(P_{Z_0 Z_1}, Q_{Z_0 Z_1}) + \text{MAE}(P_{X_0 X_1}, Q_{X_0 X_1}) + \\
    & \text{MAE}(P_{Y_0 Y_1}, Q_{Y_0 Y_1}) \Big)
\end{align}
where the subscripts attached to $P$ and $Q$ refer to the individual Pauli terms in the effective Hamiltonian $H_\text{eff}$.

\begin{table*}[h]
\centering

\begin{minipage}{0.48\linewidth}
    \raggedright
    \textbf{(i)}\\[0.2cm]
    
    \begin{tabular}{c|c}
    \hline
     & \textbf{$E_0$} \\ \hline
    IQM Garnet 20250901 & $-0.9993276$ \\
    GTH noise model $\mathcal{N}^{(2)}_{\Lambda_{q_i,q_j}}$ & $-1.0006050$ \\ \hline
    $\Delta E_0$ & 0.128\% \\ \hline
    \end{tabular}
    
    \label{table.UCC_results_20250901}
\end{minipage}
\hfill
\begin{minipage}{0.48\linewidth}
    \raggedright
    \textbf{(ii)}\\[0.2cm]
    
    \begin{tabular}{c|c}
    \hline
     & \textbf{$E_0$} \\ \hline
    IQM Garnet 20250114 & $-0.9965888$ \\
    GTH noise model $\mathcal{N}^{(2)}_{\Lambda_{q_i,q_j}}$ & $-0.9935546$ \\ \hline
    $\Delta E_0$ & 0.304\% \\ \hline
    \end{tabular}
    
    \label{table.UCC_results_20250114}
\end{minipage}

\caption{Tables (i) and (ii) show the values of IQM Garnet and the corresponding GTH noise models for successive and delayed execution of the 1-qubit GST, 2-qubit GST, and UCC sweeps. For each execution type, the expectation value $E_0$ was evaluated at $\theta = \theta^*$, and the percentage relative differences $\Delta E_0$ were computed.}
\label{table.UCC_side_by_side}
\end{table*}

\begin{figure}[h]
    \centering
    \begin{minipage}{0.47\linewidth}
        \raggedright
        \textbf{(i)}\\[0.2cm]
        \includegraphics[width=0.85\linewidth]{fig3_20250901.pdf}
    \end{minipage}
    \hfill
    \begin{minipage}{0.47\linewidth}
        \raggedright
        \textbf{(ii)}\\[0.2cm]
        \includegraphics[width=0.85\linewidth]{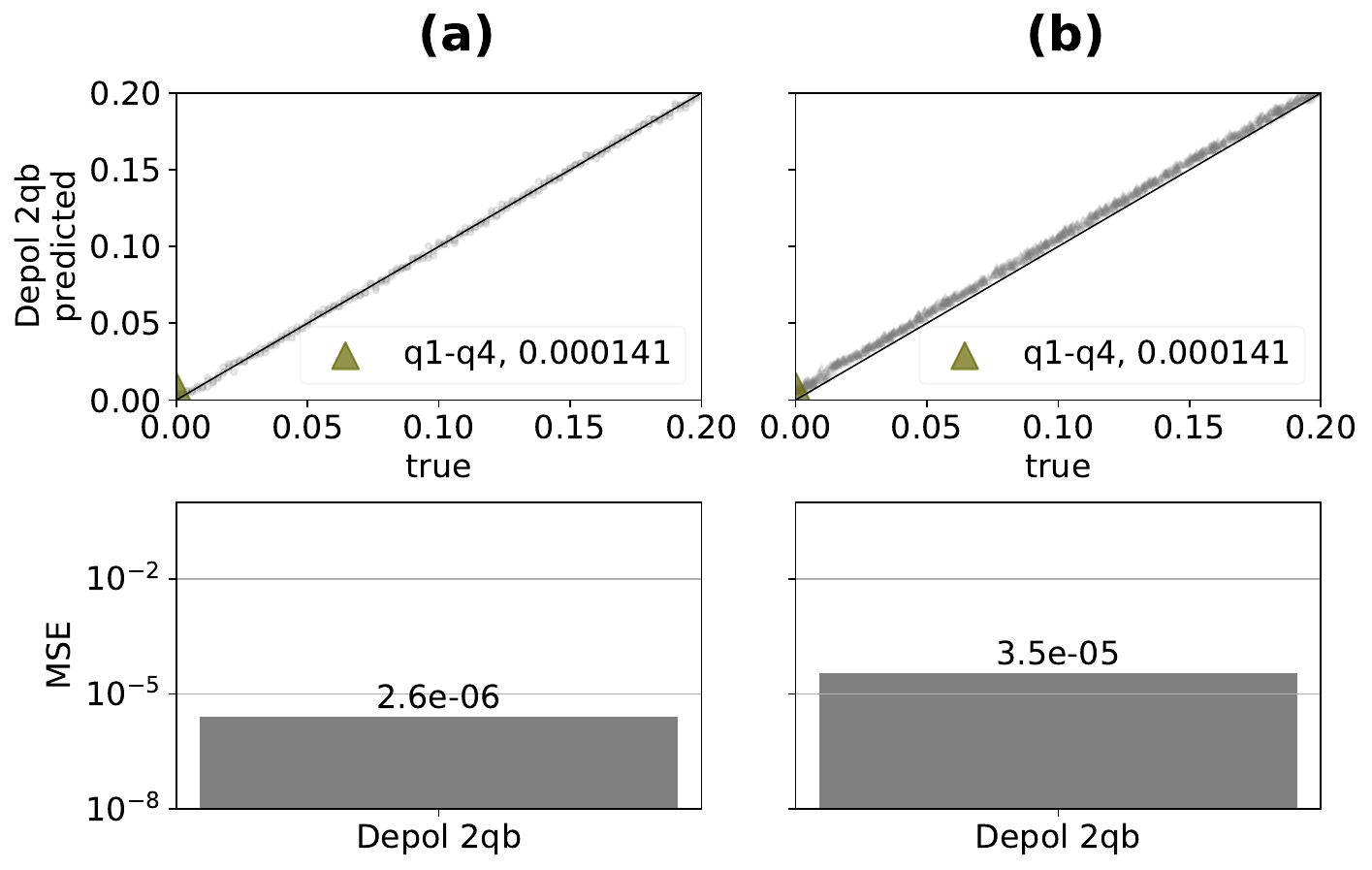}
    \end{minipage}
    \caption{Following the workflow in Fig.~\ref{fig.workflow}, every new prediction derived from 1-qubit GST on hardware requires retraining the NN-2Q. Here, we plot the training evaluations for NN-2Qs from (i) successive execution, and (ii) delayed execution, of 1-qubit GST, 2-qubit GST, and UCC sweep.
    }
    \label{fig.NN2Q_side_by_side}
\end{figure}

\begin{figure}
    \parbox{\linewidth}{\raggedright
        \textbf{(i)}\\
        \includegraphics[width=0.95\linewidth]{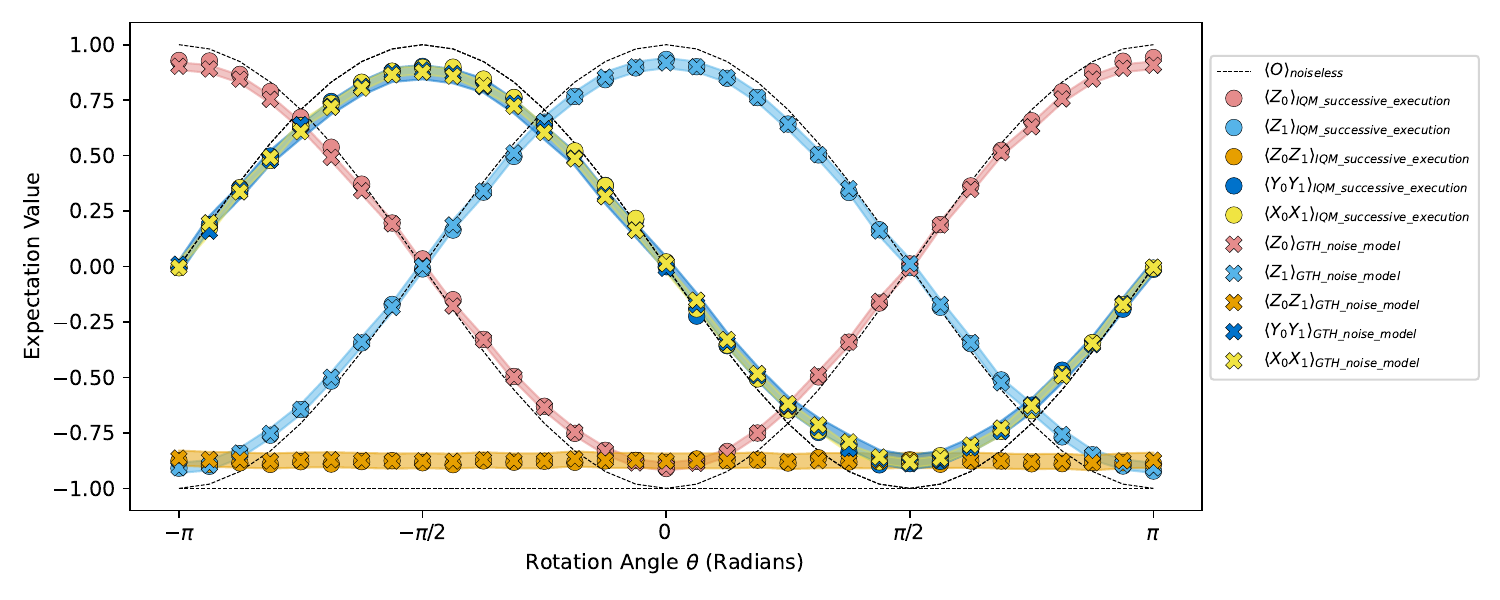} \\
        \textbf{(ii)}\\
        \includegraphics[width=0.95\linewidth]{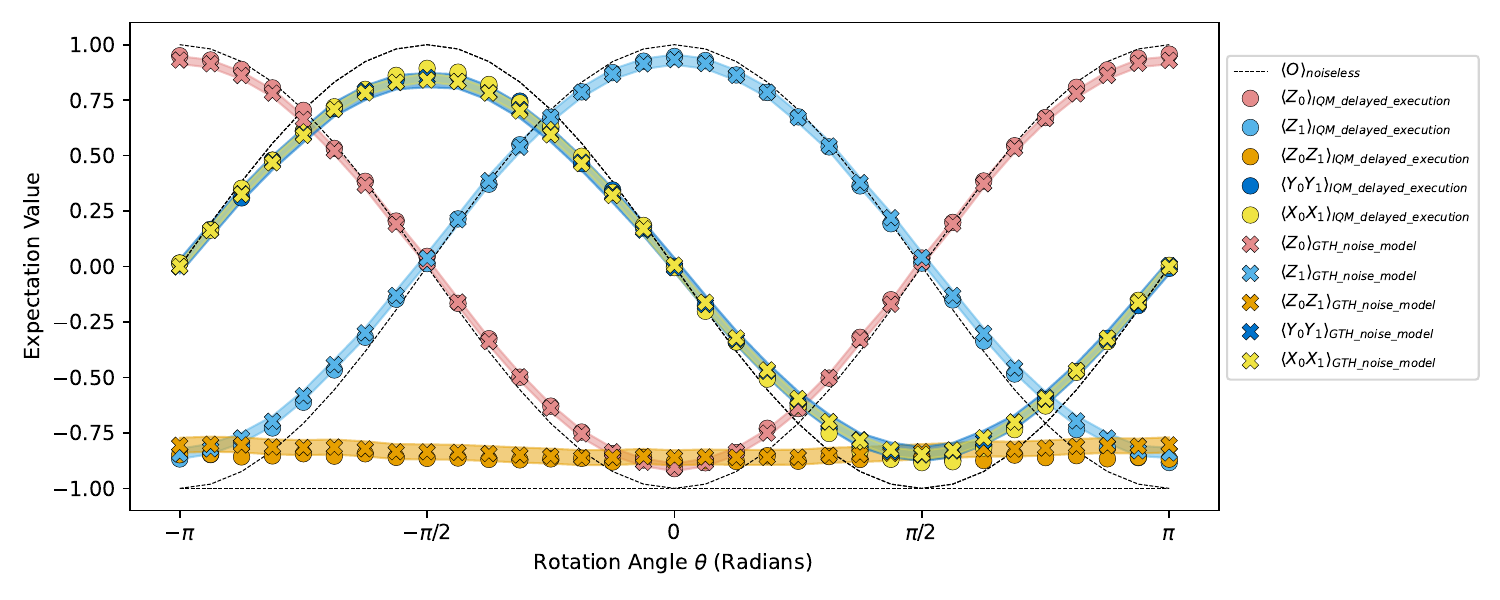} \\
    }
    \caption{The UCC sweeps for (i) successive execution (same as in main text) and (ii) delayed execution of 1-qubit GST, 2-qubit GST, and UCC sweep. The mean absolute difference for the successive execution is 0.01385 while the mean absolute difference for the delayed execution is 0.02168.} 
    \label{fig.UCC_sweep_side_by_side}
\end{figure}

\clearpage

\section{Gate set tomography}
\label{sec.appendix_GST}

\begin{figure}[h]
    \parbox{\linewidth}{\raggedright
        \textbf{(a)}\\
        \includegraphics[width=1.0\linewidth]{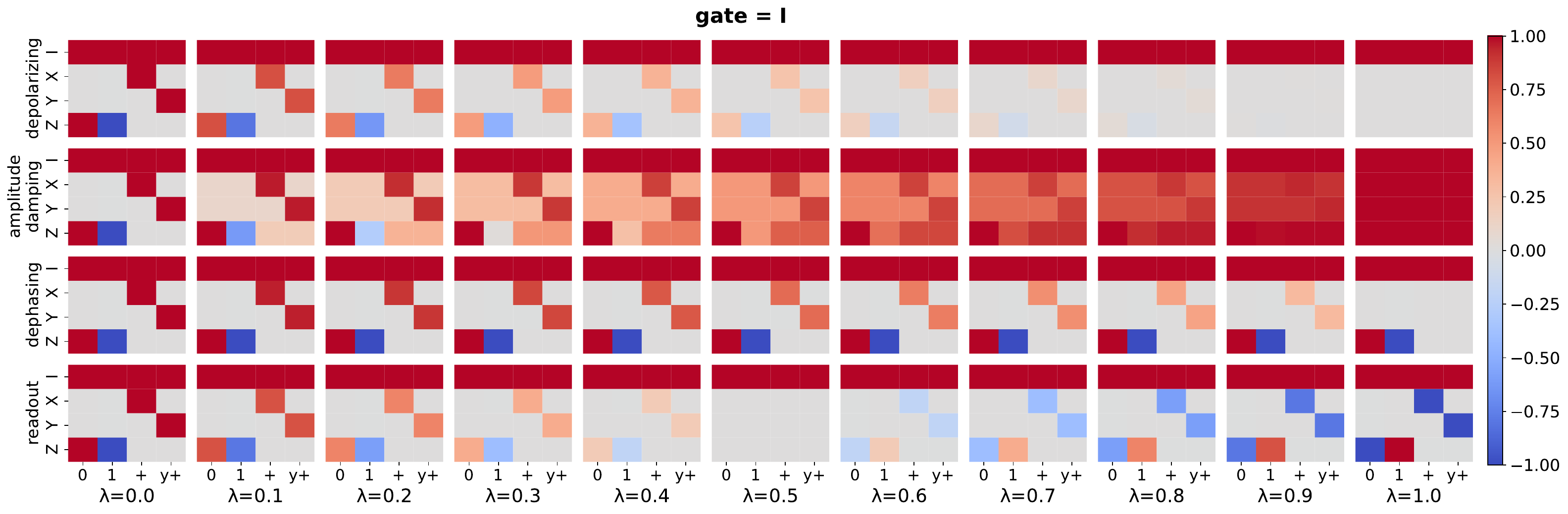} \\
        \textbf{(b)}\\
        \includegraphics[width=1.0\linewidth]{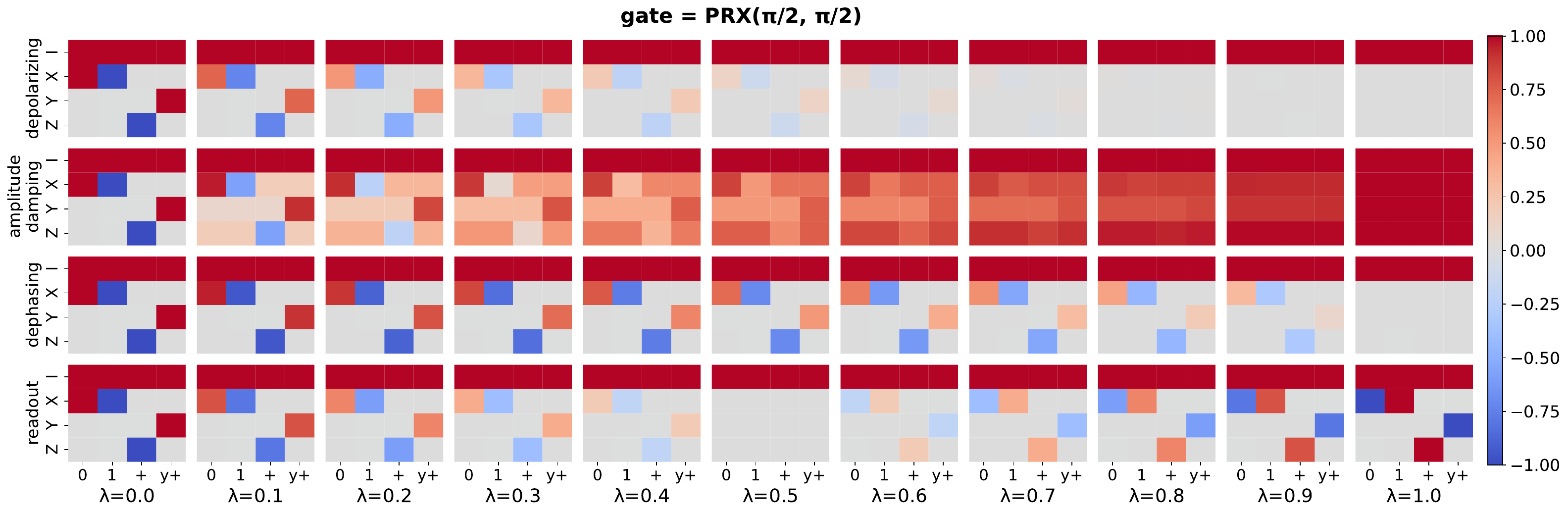}
    }
    \caption{To illustrate the effects of different noise models on different single-qubit gates, a heatmap containing the 4 by 4 matrices for the (a) identity gate and (b) $\text{PRx}(\pi/2, \pi/2)$ are shown.}
    \label{fig.GST_noise_models}
\end{figure}

\subsection{Gate set tomography preliminaries}

Recall that gate set tomography consists of state preparation, (optional addition of gate) and measuring the expectation values of Pauli operators.

The states prepared are $\rho_k \in \{\ket{0}\bra{0}, \ket{1}\bra{1}, \ket{+}\bra{+}, \ket{y+}\bra{y+} \}^{\otimes n}$, where $n$ is the number of qubits in the system. Then an $n$-qubit gate may or may not be applied after the prepared state. Finally, the expectation values of the Pauli operators are then computed, $M_j = \{\langle I \rangle, \langle X \rangle, \langle Y \rangle, \langle Z \rangle\}^{\otimes n}$.

When no gate is applied after state preparation, the resulting matrix $g$ has the following elements
\begin{equation}
\label{eq.appendix_GST_g}
    g_{jk} = \text{Tr}(M_j \rho_k).
\end{equation}
$g$ may also be regarded as a calibration matrix.

If a gate $U$ is applied after state preparation and before measuring the expectation values, the resulting matrix $\mathcal{U}$ contains the following elements
\begin{equation}
\label{eq.appendix_GST_U}
    \mathcal{U}_{jk} = \text{Tr}(M_j U \rho_k).
\end{equation}

For single-qubit GST ($n=1$), the resultant matrices $g$ and $\mathcal{U}$ are size 4 $\times$ 4. For two-qubit GST ($n=2$), the resultant matrices $g$ and $\mathcal{U}$ are size 16 $\times$ 16.

\subsection{Effects of different noise models on GST matrices}

In the remainder of this section, we will focus on single-qubit GST of two gates under the effects of different noise models, heuristically demonstrating the importance of carefully selecting gates for the GTH noise model. The two gates we use in our example are the identity gate and the $\text{PRx}(\theta=\frac{\pi}{2}, \phi=\frac{\pi}{2})$ gate.

Since the first row of the GST matrices $g$ and $\mathcal{U}$ corresponds to the expectation value of the identity operator, which is computed by the sum of all probabilities, these elements remain invariant at 1.0 regardless of noise. 

As shown in Fig. \ref{fig.GST_noise_models}, both the identity gate and $\text{PRx}(\theta=\frac{\pi}{2}, \phi=\frac{\pi}{2})$ gate undergo similar transformations when depolarizing noise is increased from $\lambda=0$ to $1$, and when readout noise is increased from $\lambda=0$ to $\lambda=0.5$. This similarity highlights cases where distinct noise processes may generate overlapping features in the GST representation.

Focusing on the $\text{PRx}(\theta=\frac{\pi}{2}, \phi=\frac{\pi}{2})$ gate, we observe that dephasing noise at $\lambda \approx 0.60$ can be mistaken as depolarizing noise at $\lambda \approx 0.30$. Such overlaps indicate that different noise models can, under certain conditions, yield GST matrices with comparable signatures.

At the same time, some gates, such as the identity gate, have matrix elements other than the expectation values of the identity operator that remain robust under dephasing noise but not depolarizing noise. Building on this observation, we selected a variety of angles for $\theta$ and $\phi$ such that the resulting 4 by 4 single-qubit GST matrices show a clear distinction between depolarizing noise and dephasing noise, as detailed in Table~\ref{table.PRX_gate_set}. This design choice provides the neural network with sharper discriminative features, enabling more reliable differentiation between noise types.

\subsection{Time and costs involved}
\label{sec.appendix_GST_time_costs}

The total cost for the single-qubit GST (with our chosen gate set in Table~\ref{table.PRX_gate_set}) can be estimated by finding the total number of single-qubit circuits and two-qubit circuits. With 15 PRx gates and an empty single-qubit $g$ matrix, we have a total of 16 single-qubit matrices (each with 16 circuits), bringing the total to at most $15 \times 16 = 240$ circuits. The two-qubit GST is made up of one CZ gate and an empty single-qubit $g$ matrix (each with $240$ circuits), thus requiring at most $2 \times 256 = 512$ circuits.

In our methodology, we do $10,000$ shots for single-qubit GST and $1,000$ shots for two-qubit GST. Referencing the circuit execution costs given in Appendix~\ref{sec.appendix_garnet}, we can estimate that the total cost for a full run of GST is approximately $240 \times (0.30 + 0.00145 \times 10,000) + 512 \times (0.30 + 0.00145 \times 1,000) \approx 4,100$ USD on AWS Braket.
With $10,000$ shots for each circuit in the Quantum Chemistry example, the cost is estimated to be approximately $1,500$ USD on AWS Braket.

Next we estimate the feasibility of running only the GST portion of our protocol with the 10-minutes 
free Open Plan of IBM Quantum. Note that IBM Quantum employs a Fair share scheduler~\cite{FairShareScheduler} to adjust users' priority. Priority is decreased as the number of jobs executed on the backends is increased. Consider the scenario where the diminishing priority from the Fair share scheduler is ignored, where one only needs to be mindful of the free-access time limit. Running $240 + 512 = 752$ circuits demands that each circuit returns the measurement outcomes in at most $(10 \text{min} \times 60 \text{sec}) / 752 = 0.7979$ secs. Therefore, it is highly infeasible to run the protocol on the 
free Open Plan especially with the inclusion of the Fair share scheduler.

\section{Additional information to construct the GTH noise model $\mathcal{N}^{(2)}_{\Lambda_{q_i, q_j}}$}
\label{sec.appendix_recipe}

This section provides additional information for Section~\ref{sec.recipe} in the main text where we listed explicitly the steps to construct the GTH noise model. Both single-qubit and two-qubit \textit{simulated} GST training data were obtained with GST, each performed with 10,000 shots. The parameter values for $\Lambda$ were sampled in the range $0 \leq \lambda_d, \lambda_a, \lambda_f, \lambda_r < 0.1$ using discrete, uniform steps of 0.01.

The choice of uniform step size for $\Lambda$ is crucial in balancing training data size and sampling resolution. While a smaller step size gives finer granularity of data for the neural network, it inevitably incurs a longer training time and necessitates a more complex model. Since $\Lambda := (\lambda_d, \lambda_a, \lambda_f, \lambda_r)$ has four components, the dataset size grows exponentially with finer granularity. Through experimentation, we found that a uniform step size of 0.01 for $\Lambda$ provides the best balance between accuracy and computational cost. 

Similarly, the values for $\zeta$ were sampled in the range $0 \leq \zeta < 0.2$ in discrete, uniform steps of 0.002. Unlike $\Lambda$, the step size for $\zeta$ was chosen to be finer because the data size scales only linearly with one component, making it computationally feasible to use a higher resolution. Due to the small step size, the inaccuracy in extrapolation errors for $\zeta > 0.1998$ are negligible in Fig.~\ref{fig.NN-2Q_results}. 

To prevent statistical fluctuations in the GST matrices from being misinterpreted as hardware noise, 9 GST runs were conducted for each $\Lambda$ value, and 50 for each $\zeta$ value, mitigating the effects of finite-shot noise.

\section{Details of IQM Garnet device}
\label{sec.appendix_garnet}

The IQM Garnet device is available on Amazon Braket as a pay-per-use device. Its technology is based on superconducting transmon qubits, featuring a 20-qubit chip with a lattice-like topology shown in Fig.~\ref{fig:garnet}. The full technical details of the IQM Garnet chip can be found in Ref. \cite{abdurakhimov2024technology}.
\begin{figure}[h]
    \centering
    \includegraphics[width=0.4\linewidth]{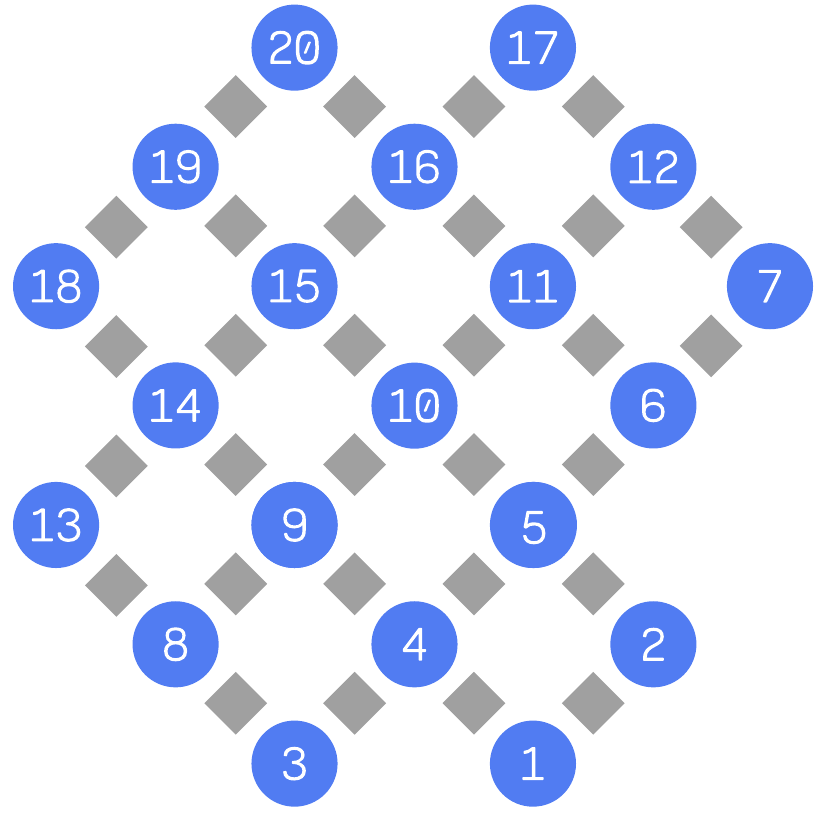}
    \caption{The topology of IQM Garnet device. Qubits are labeled from 1 to 20 in a square-like lattice.}
    \label{fig:garnet}
\end{figure}

The native gates of IQM Garnet are the single-qubit Phase Rx gate (PRx) and the two-qubit CZ gate. The Phase Rx gate contains two angles and can be decomposed into three rotations:
\begin{equation}
    \text{PRx}(\theta, \phi) = \text{Rz}(\phi) \text{Rx}(\theta) \text{Rz}(-\phi)
\end{equation}

To execute Qibo circuits on the IQM Garnet chip, we use the \texttt{BraketClientBackend}, that is found on \texttt{qibo-cloud-backends}, to execute quantum circuits. Note that one needs to write the quantum circuits with native gates together with verbatim mode in order avoid any on-device transpilation. This will force the execution of circuits on the specified qubits as well.

IQM makes available the Garnet chip's calibration data on Amazon Braket. Below, we show two code blocks, one for the single-qubit properties and the other for two-qubit properties, that illustrate how the calibration data looks like.

\begin{lstlisting}[language=Python, caption=Example of single-qubit calibration data for qubit 1. We were made aware that Amazon Braket has since included the probabilities of readout errors (0 to 1 and 1 to 0) in addition to the readout fidelity some time after our work.]
"oneQubitProperties": {
    "1": {
      "T1": {
        "value": 0.000041987029524848376,
        "standardError": 0.000002087249176131054,
        "unit": "S"
      },
      "T2": {
        "value": 0.000009097040676807278,
        "standardError": 5.706318009871576e-7,
        "unit": "S"
      },
      "oneQubitFidelity": [
        {
          "fidelityType": {"name": "SIMULTANEOUS_RANDOMIZED_BENCHMARKING"},
          "fidelity": 0.9992841052618756,
          "standardError": 0.000025954746699487628
        },
        {
          "fidelityType": {"name": "READOUT"},
          "fidelity": 0.98175
        }
      ]
    },
\end{lstlisting}

\begin{lstlisting}[language=Python, caption=Example of two-qubit calibration data for coupled qubits 2-5.]
"twoQubitProperties": {
    "2-5": {
      "twoQubitGateFidelity": [
        {
          "gateName": "CZ",
          "fidelity": 0.9889978279806257,
          "standardError": 0.0008092654295870103,
          "fidelityType": {"name": "SIMULTANEOUS_INTERLEAVED_RANDOMIZED_BENCHMARKING"}
        },
        {
          "gateName": "Two_Qubit_Clifford",
          "fidelity": 0.9759941470601171,
          "standardError": 0.0004535443935232237,
          "fidelityType": {"name": SIMULTANEOUS_INTERLEAVED_RANDOMIZED_BENCHMARKING"}
        }
      ]
    },
\end{lstlisting}

\section{Comparison between GTH model and IBMQ noise model}\label{sec.appendix_gth_vs_ibmqnoisemodel}

In this section, we assess the performance of the IBMQNoiseModel~\cite{IBMQNoiseModel}, a noise model that requires device calibration data inputs such as T1, T2 times in order to implement the depolarizing channel, thermal relaxation channel, and the readout error channel.

The IQM Garnet's calibration data for 2025-09-01 was pulled from the Amazon Braket server and used as inputs to the IBMQNoiseModel. As the excited population values are not published, we set it to 0 following Ref.~\cite{georgopoulos2021modeling}.

The UCC sweeps for the IBMQNoiseModel at different excited populations are plotted in Fig.~\ref{fig.IBMQNoiseModel_ep}. The mean absolute difference for the IBMQNoiseModel is found to be 0.0196.

\begin{figure}
    \centering
    \includegraphics[width=1.0\linewidth]{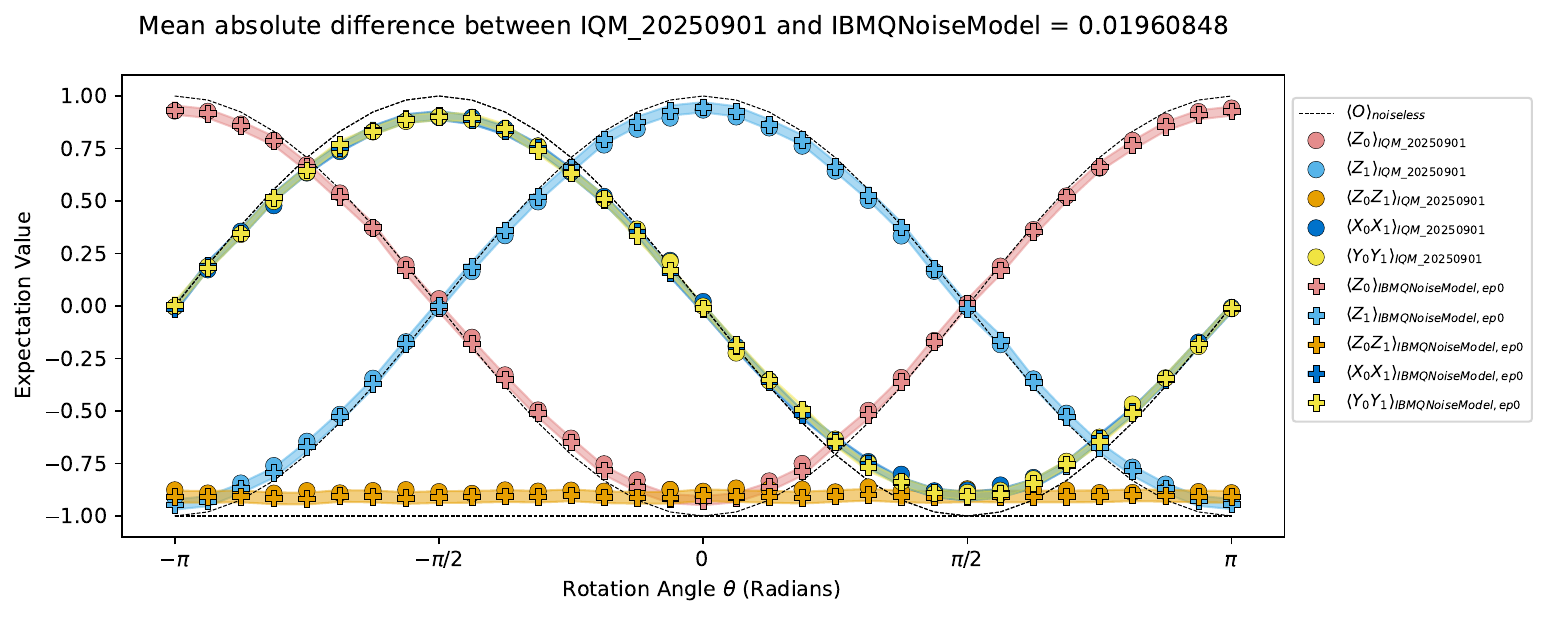}
    \caption{The UCC sweep for IBMQNoiseModel with mean absolute difference of 0.0196.}
    \label{fig.IBMQNoiseModel_ep}
\end{figure}

\end{widetext}
\end{document}